\documentclass{aa}
\usepackage{lscape}
\usepackage[varg]{txfonts}
\usepackage[dvips]{graphicx}
\usepackage{xspace}
\usepackage{subfigure}
\usepackage{comment}
\usepackage{natbib, aas_macros}
\begin{document}
\title{Optical -- Near-Infrared catalogue for the $AKARI$ North Ecliptic Pole Deep Field.}
\author{Nagisa Oi
  \inst{1}
  \and Hideo Matsuhara\inst{1}
  \and Kazumi Murata\inst{1,2}
  \and Tomotsugu Goto\inst{3}
  \and Takehiko Wada\inst{1}
  \and Toshinobu Takagi\inst{1}
  \and Youichi Ohyama\inst{4}
  \and Matthew Malkan\inst{5}
  \and Myungshin Im\inst{6} 
  \and Hyunjin Shim\inst{6, 7}
  \and Stephen Serjeant\inst{8}
  \and Chris Pearson\inst{8, 9, 10}
}

\institute{
  Institute of Space and Astronautical Science, Japan Aerospace Exploration Agency, Sagamihara, 252-5210 Kanagawa, Japan\\
\email{nagisaoi@ir.isas.jaxa.jp} 
\and
Department of Space and Astronautical Science, The Graduate University for Advanced Studies, Japan
\and
Institute of Astronomy and Department of Physics, National Tsing Hua University No. 101, Section 2, Kuang-Fu Road, Hsinchu 30013, Taiwan, R.O.C
\and
Academia Sinica, Institute of Astronomy and Astrophysics, 11F of Astronomy-Mathematics Building, National Taiwan University, No.1, Sec. 4, Roosevelt Rd, Taipei 10617, Taiwan R.O.C.
\and 
Department of Physics and Astronomy, UCLA, Los Angeles, CA, 90095-1547, USA
\and 
Department of Physics \& Astronomy, FPRD, Seoul National University, Shillim-Dong, Kwanak-Gu, Seoul 151-742, Korea
\and
Department of Earth Science Education, Kyungpook National University, Daegu 702-701, Republic of Korea
\and 
Astrophysics Group, Department of Physics, The Open University, Milton Keynes, MK7 6AA, UK
\and 
RAL Space, Rutherford Appleton Laboratory, Harwell Oxford, OX11 0QZ, UK
\and
Oxford Astrophysics, Oxford University, Keble Road, Oxford OX1 3RH, UK
}

\date{Received \today; accepted xxx xx, xxxx}

\abstract
{}
{
We present an 8-band ($u^{*}$, $g^{'}$, $r^{'}$, $i^{'}$, $z^{'}$, $Y$, $J$, $K_{\rm s}$) optical to near-infrared deep photometric catalogue based on the observations made with MegaCam and WIRCam at CFHT, and compute photometric redshifts, $z_{\rm p}$ in the North Ecliptic Pole (NEP) region.
$AKARI$ infrared satellite carried out deep survey in the NEP region at  near to mid infrared wavelength.
Our optical to near-infrared catalogue provides us to identify the counterparts, and $z_{\rm p}$ for the $AKARI$ sources.
}
{We obtained seven ($g^{'}$, $r^{'}$, $i^{'}$, $z^{'}$, $Y$, $J$, $K_{\rm s}$) band imaging data, and we cross-matched them with existing $u^{*}$-band data (limiting magnitude = 24.6 mag [5$\sigma$; AB]) to design the band-merged catalogue.
We included all $z^{'}$-band  sources with counterparts in at least one of the other bands in the catalogue.
We used a template-fitting methods to compute $z_{\rm p}$ for all the catalogued sources. 
}
{
The estimated 4$\sigma$ detection limits within an 1 arcsec aperture radius are 26.7, 25.9, 25.1, and 24.1 mag [AB] for the optical $g^{'}$, $r^{'}$, $i^{'}$, and $z^{'}$-bands and 23.4, 23.0, and 22.7 mag for the near-infrared $Y$, $J$, and $K_{\rm s}$-bands, respectively.
There are a total of 85$\hspace{0.1em}$797 sources in the band-merged catalogue.
An astrometric accuracy of this catalogue determined by examining coordinate offsets with regard to 2MASS is 0.013 arcsec with a root mean square (RMS) offset of 0.32 arcsec. 
We distinguish 5441 secure stars from extended sources using the $u^{*}-J$ versus $g^{'}-K_{\rm s}$ colours, combined with the SExtractor stellarity index of the images. 
Comparing with galaxy spectroscopic redshifts, we find a photometric redshift dispersion, $\sigma_{\Delta z/(1+z)}$, of 0.032 and catastrophic failure rate, $\frac{\Delta z}{1+z}>0.15$, of 5.8\% at $z<1$, while a dispersion of 0.117 and a catastrophic failures rate of 16.6\% at $z>1$.
We extend estimate of the $z_{\rm p}$ uncertainty over the full magnitude/redshift space with a redshift probability distribution function and find that our redshift are highly accurate with $z^{'}<22$ at $z_{\rm p}<2.5$ and for fainter sources with $z^{'}<24$ at $z<1$.
From the investigation of photometric properties of $AKARI$ infrared sources (23$\hspace{0.1em}$354 sources) using the $g^{'}z^{'}K_{\rm s}$ diagram, $<$ 5\% of $AKARI$ sources with optical counterparts are classified as high-$z$ ($1.4<z<2.5$) star-forming galaxies. Among the high-$z$ star-forming galaxies, $AKARI$ mid-infrared detected sources seem to be affected by stronger dust extinction compared with sources with non-detections in the AKARI MIR bands. 
The full, electronic version of our 8-band merged catalogue with $z_{\rm p}$ will be available at the CDS.
}
    {}

\titlerunning{Optical and Near-Infrared merged catalogue at $AKARI$ NEP-Deep field}
    \maketitle
\section{INTRODUCTION}
An understanding of the cosmic history of star formation is one of the most important aspects in the study of galaxies.
It has been shown that the star formation rate density derived from UV and optical wavelengths at $z\sim1$ is one order of magnitude higher than in the local Universe \citep{1995ApJ...441...18M, 2001AJ....122..288H, 1997ApJ...486L..11C, 1999ApJ...519....1S, 2007ApJ...657..738L}. 
These results highlight the importance of studying the cosmic star formation activity at high redshift ($z\sim1$).
Studies of the extragalactic background have suggested at least one third (or half) of the luminous energy generated by stars has been reprocessed into the infrared by dust \citep{1999A&A...344..322L, 1996A&A...308L...5P, 2006ApJ...648..774S, 2008A&A...487..837F}, and that the contribution of infrared luminous sources to the cosmic infrared luminosity density increases with redshift, especially at $z>1$ \citep{2010A&A...514A...6G}.
For luminous infrared galaxies 
($L_{\rm IR}\hspace{0.3em}\raisebox{0.4ex}{$>$}\hspace{-0.65em}\raisebox{-.7ex}{$\sim$}\hspace{0.3em} 10^{11}L_{\odot}$), the SFR derived from UV continuum or optical diagnostics (H$\alpha$ emission line) is smaller than that inferred from the infrared or radio luminosity, which are free from dust extinction.
The tendency becomes more conspicuous with increasing infrared luminosity.  
This suggests that more luminous galaxies, i.e., galaxies with larger SFR, are more heavily obscured by dust \citep{2001AJ....122..288H, 2001ApJ...558...72S, 2002ApJ...581..205H, 2006ApJ...637..227C, 2007ApJS..173..404B}.
Therefore, to understand the star-formation history, it is essential to investigate star-formation in highly dust-obscured galaxies at high redshift ($z$\hspace{0.3em}\raisebox{0.4ex}{$>$}\hspace{-0.65em}\raisebox{-.7ex}{$\sim$}\hspace{0.3em}1) in the mid-infrared region where dust extinction is less severe.

The infrared space telescope $AKARI$ \citep{2007PASJ...59S.369M} was launched in February 2006 and the $AKARI$/$IRC$ \cite[InfraRed Camera: ][]{2007PASJ...59S.401O} obtained higher quality near -- mid infrared data than the previous infrared space missions such as Infrared Astronomical Satellite \citep{1984ApJ...278L...1N} and Infrared Space Observatory \citep{1996A&A...315L..27K}.
The $IRC$ has three channels (NIR, MIR-S, and MIR-L) with nine filters ($N2, N3, N4, S7, S9W, S11, L15, L18W$, and $L24$), and  $AKARI$ carried out a deep survey program in the direction of North Ecliptic Pole (NEP), so-called $NEP$-$Deep$ $survey$ \citep[hereafter, NEP-Deep: ][]{2006PASJ...58..673M} with all the nine $IRC$ bands.
Each single band observed $\sim$0.6 square degrees, and $\sim$ 0.5 square degrees circular area centered on $\alpha$ = 17$^h$55$^m$24$^s$, $\delta$ = $+$66$^{\circ}$37$^{'}$32$^{''}$ are covered by all the nine bands.
The full width at half maximum (FWHM) of each IRC channel is $\sim$ 4 arcsec in NIR, $\sim$5 arcsec in MIR-S, and $\sim$6 arcsec in MIR-L.
\cite{2013A&A...559A.132M} constructed a revised near to mid infrared catalogue of the $AKARI$ NEP-Deep.
The 5$\sigma$ detection limits in the $AKARI$ NEP-Deep catalogue are 13, 10, 12, 34, 38, 64, 98, 105 and 266 $\mu$Jy at the $N2, N3, N4, S7, S9W, S11, L15, L18W$, and $L24$ bands, respectively.
The real strength of the NEP-Deep survey lies in the unprecedented photometric coverage from 2 $\mu$m to 24 $\mu$m, critically including the wavelength domain between Spitzer$^{'}$s $IRAC$ \citep{2004ApJS..154...10F} and MIPS \citep{2004ApJS..154...25R} instruments from 8 $\mu$m to 24 $\mu$m, where only limited coverage is available from the peak-up camera on the IRS \citep{2004ApJS..154...18H}.
Therefore, NEP-Deep survey has a great advantage to be able to study infrared galaxies without uncertainties of interpolation over the mid-infrared wavelength range.

Furthermore, the NEP-Deep field is a unique region because many astronomical satellites have accumulated many deep exposures covering this location due to the nature of its position on the sky. 
X-ray observations by $Chandra$, UV observations by $GALEX$, and far-infrared observations by $Herschel$ were made toward the NEP-Deep field. 
In addition, sub-mm data from $SCUBA$-2 on the James Clerk Maxwell Telescope ($JCMT$) and radio data from Westerbork Synthesis Radio Telescope ($WSRT$) were also obtained in the NEP-Deep field \citep[see also][] {2008PASJ...60S.517W}. 
Thus this multi-band $AKARI$ NEP-Deep field data set provides a unique opportunity to study star formation of infrared luminous galaxies. 

Deep ancillary optical and near-infrared data covering the entire NEP-Deep region are essential for finding counterparts of sources detected with $AKARI$.
In addition, multi-wavelength data are needed for accurate photometric redshift estimation, especially in covering the rest-frame wavelength range 
around the 4000$\AA$ break. 
Many studies have been made of the optical counterparts of the $AKARI$ sources.
\cite{2012A&A...537A..24T} used the Subaru/Suprime cam (S-cam) to obtain deep optical images (28.4 mag for $B$-band, 28.0 mag for $V$, 27.4 mag for $R$, 27.0 mag for $i^{'}$, and 26.2 mag for $z^{'}$ in the AB magnitude system), but these data only cover one half of the NEP-Deep area. 
\cite{2007ApJS..172..583H} observed the entire NEP-Deep field with Canada France Hawaii Telescope (CFHT) in the $g^{'}$, $r^{'}$, $i^{'}$, and $z^{'}$-bands with limiting magnitudes of 26.1, 25.6, 24.7, and 24.0 mag (4$\sigma$) with 1 arcsec diameter apertures, however these images have insensitive areas resulting from gaps between CCD chips. 
\cite{2007AJ....133.2418I} observed the area with KPNO-2.1 m/FLAMINGOS in the $J$- and $K_{\rm s}$-bands, but the 3$\sigma$ detection limits were shallow 
(21.85 mag and 20.15 mag in the $J$ and $K_{\rm s}$-bands, respectively). Moreover, these data cover only about one third of the NEP-Deep field ($25^{'}\times30^{'}$). 

\begin{figure}[t]
\begin{center}
  \includegraphics[width=90mm]{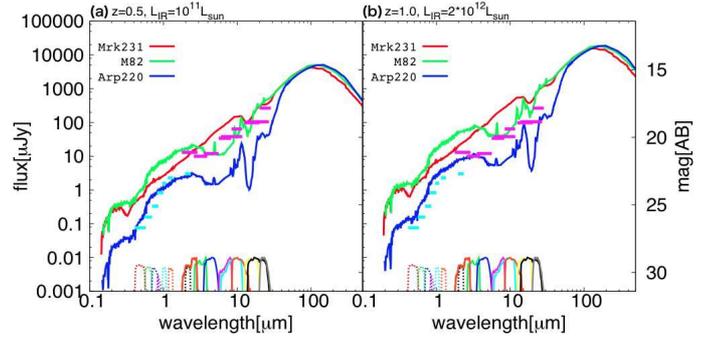}
  \caption{Example of galaxies SEDs. Left panel shows the SEDs with redshift of 0.5 with infrared luminosity of 10$^{11}L_{\odot}$, while right panel is for the SEDs with redshift of 1.0 with infrared luminosity of 2$\times$10$^{12}L_{\odot}$. Red, Green and Blue solid lines represent SEDs of Mrk231, M82, and Arp220. Thick horizontal lines in pink and sky blue show the 5$\sigma$ detection limits of $AKARI$ $N2$ -- $L24$ filters and ground based $g^{'}$ -- $K_{\rm s}$ band 4 $\sigma$ limiting magnitudes which we newly observed (see Table \ref{tb:photometry}).  Filter response curves of CFHT/MegaCam, WIRCam, and $AKARI/IRC$ are shown at the bottom of the panels. Red, green, blue, pink, cyan, orange, and black dotted-lines represent $g^{'}$, $r^{'}$, $i^{'}$, $z^{'}$, $Y$, $J$, and $K_{\rm s}$, while red, green, blue, pink, cyan, yellow, black, orange, and gray solid lines are $AKARI/IRC$ $N2$ -- $L24$ bands, respectively.}
  \label{fig:SEDexampleOptAKARI}
\end{center}
  \end{figure}

\begin{figure}[htbp]
\begin{center}
  \includegraphics[width=90mm]{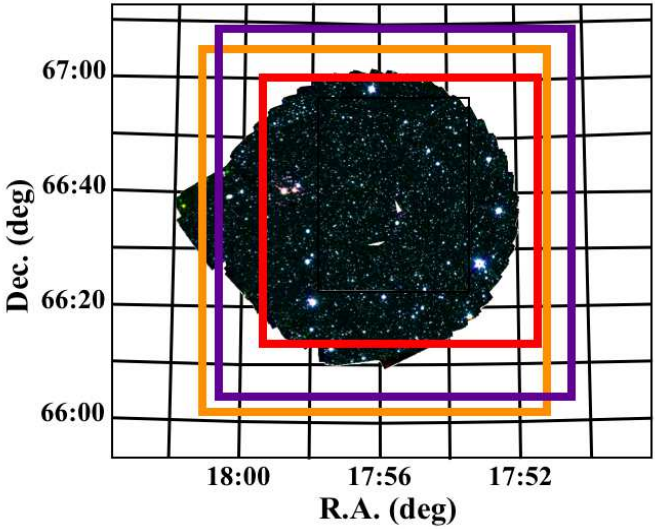}
  \caption{Coverages of various surveys in the NEP-Deep field. The colour image in the middle is the three-colour near-infrared image made with the $IRC$ data (blue--2 $\mu$m, green--3 $\mu$m, red--4 $\mu$m). The orange box is the survey coverage of MegaCam $g^{'}$, $r^{'}$, $i^{'}$, and $z^{'}$, and the red box is that of WIRCam $Y$, $J$, and $K_{\rm s}$. The purple box represents the MegaCam $u^{*}$-band observation field \citep{2012A&A...537A..24T}.}
  \label{fig:NEPD}
\end{center}
  \end{figure}

Figure \ref{fig:SEDexampleOptAKARI} is an example of SEDs for star-forming galaxies, Mrk231, M82, and Arp220 if their redshifts are at $z=0.5$ with infrared luminosity of 10$^{11}L_{\odot}$ (left) or $z=1.0$ with infrared luminosity of 2$\times$10$^{12}L_{\odot}$. In order to optically identify Arp220 at $z=1$ (blue SED) which can be detected by $AKARI$, 26 -- 23 mag [AB] for $g^{'}$--$z^{'}$ bands and 22 -- 23 mag [AB] for near-infrared bands are required.
In this study, we present deep optical $g^{'}$, $r^{'}$, $i^{'}$, and $z^{'}$ (orange square in Figure \ref{fig:NEPD}) and near infrared $Y$, $J$, and $K_{\rm s}$ imaging data (red square in Figure \ref{fig:NEPD}) in the NEP-Deep field using MegaCam \citep{2003SPIE.4841...72B} and the Wide-field InfraRed Camera \citep[WIRCam;][]{2004SPIE.5492..978P} on the CFHT.
We also combine existing $u^{*}$-band data \citep[purple square in Figure \ref{fig:NEPD}, PI: S. Serjeant; ][]{2012A&A...537A..24T} to provide a band-merged catalogue based on the $z^{'}$-band.
The catalogue will be used to analyze the $AKARI$ NEP sources.
This paper is organized as follows; we describe our observations and data reduction procedures in $\S$2, and source extraction in each band in $\S$3.
We present our band-merged catalogue in $\S$4.
In $\S$5, we describe our method of star-galaxy separation based on colour-colour-criteria and the SExtractor stellarity index.
We calculate photometric redshifts for the sources in our catalogue by Spectral Energy Distribution (SED) fitting which is presented in $\S$6. 
$\S$7 provides properties of $AKARI$ sources identified their optical counterparts in our catalogue.
We summarize our results in the final section.

\section{OBSERVATION and DATA REDUCTION}
\label{sec:OBSERVATION and DATA REDUCTION}
\subsection{CFHT/MegaCam (optical: $g^{'}$, $r^{'}$, $i^{'}$, and $z^{'}$-bands)}
We obtained optical ($g^{'}$, $r^{'}$, $i^{'}$, and $z^{'}$) imaging of nearly the entire NEP-Deep field with CFHT/MegaCam at the telescope prime focus.
The observations were carried out in queue mode, spread over 22 photometric nights from 2011 April 27 to 2011 September 4, resulting in a total of 166 frames (PI: T. Goto).
MegaCam is a wide-field imager composed of 36 2048$\times$4612 pixel CCDs, covering a full 1$\times$1 deg$^2$ field of view with a pixel scale of 0.187 arcsec. 
The MegaCam $g^{'}$, $r^{'}$, $i^{'}$, and $z^{'}$-band filters follow the original Sloan Digital Sky Survey \citep[SDSS;][]{2000AJ....120.1579Y} filters\footnote{http://www3.cadc-ccda.hia-iha.nrc-cnrc.gc.ca/megapipe/docs/filters.html}. 

\begin{figure}[htbp]
\begin{center}
\includegraphics[width=50mm]{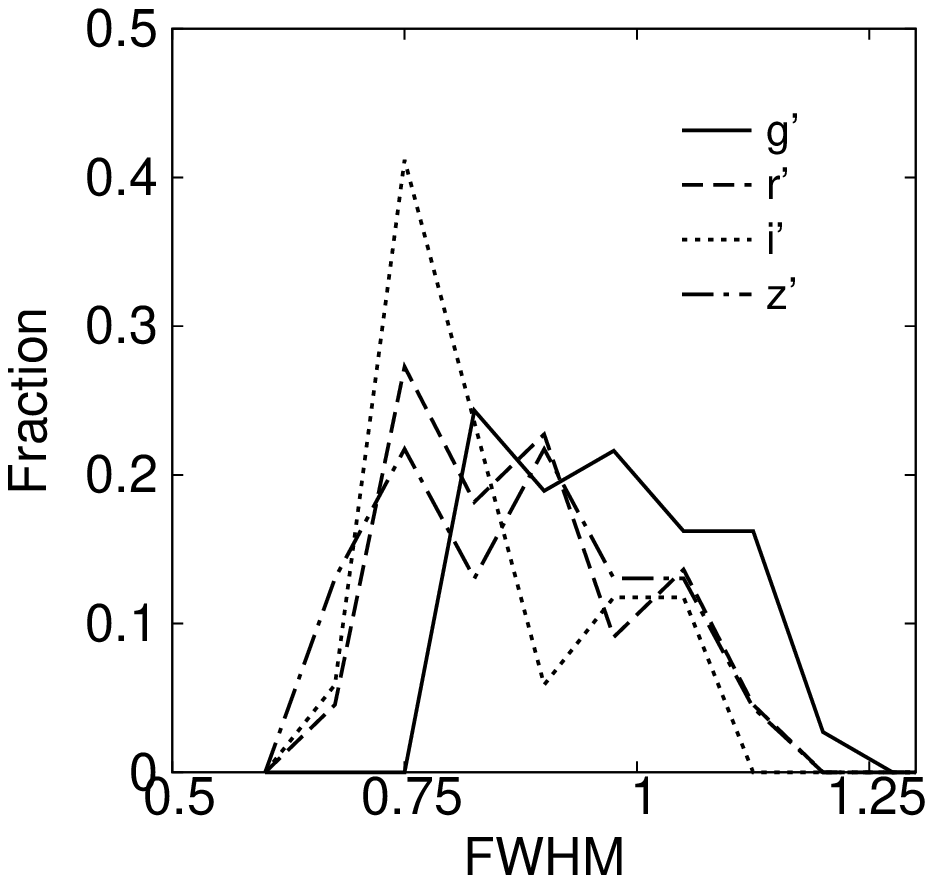}
\hspace{-1.2cm}
\includegraphics[width=50mm]{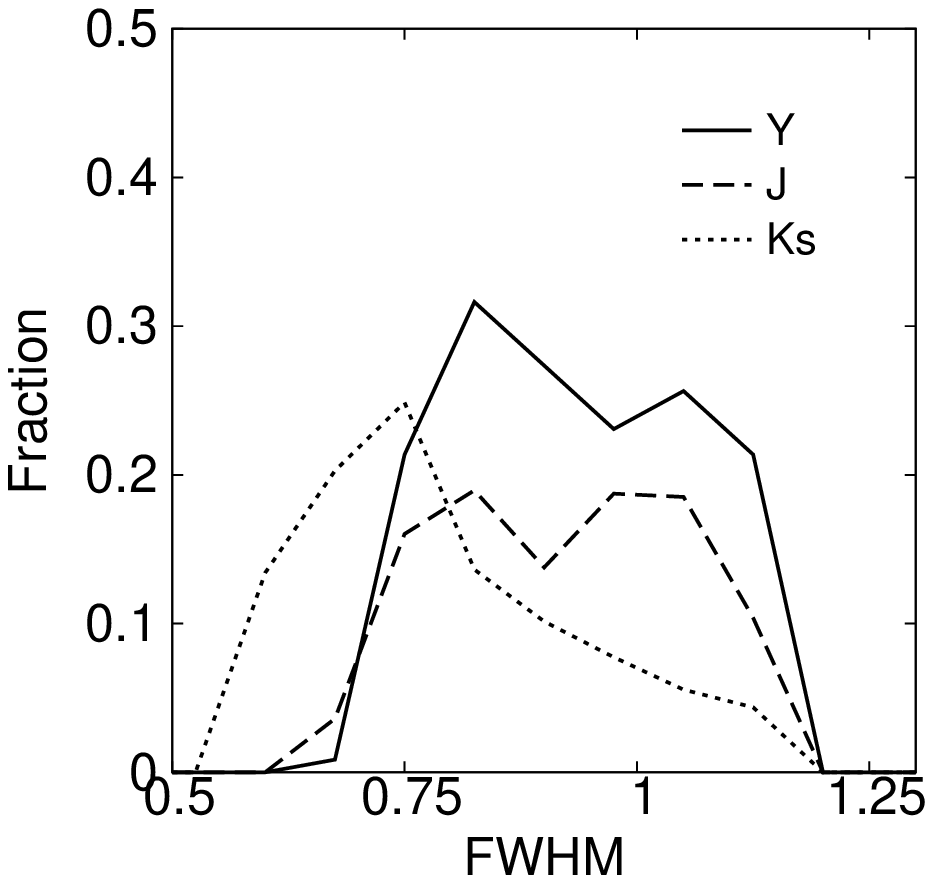}
  \caption{FWHM distribution of each flame we used in optical (left) and infrared (right). The solid, dashed, dotted, and dashed-dotted lines in the left panel show $g^{'}r^{'}i^{'}z^{'}$-bands, respectively, while the solid, dashed, and dotted lines in the right panel represent $YJK_{\rm s}$-bands.
 }
   \label{fig:FWHMDistribution}
\end{center}
  \end{figure}

The pre-processing, including bias subtraction and flat-fielding of raw images, was carried out using the Elixir pipeline \citep{2004PASP..116..449M}. Elixir is a collection of programs and databases, specializing for MegaCam data.
By stacking the Elixir-processed images, we produced a final mosaiced image using the AstrOmatic softwares including WeightWatcher\footnote{http://www.astromatic.net/software/weightwatcher} \citep{2008ASPC..394..619M} for creating a weight map, SExtractor\footnote{http://www.astromatic.net/software/sextractor} \citep{1996A&AS..117..393B} for extraction and photometry of sources, SCAMP\footnote{http://www.astromatic.net/software/scamp}\citep{2006ASPC..351..112B} for astronomical calibration with 2MASS as an astrometric reference catalogue, and modified SWarp\footnote{http://www.astromatic.net/software/swarp} \citep{2002ASPC..281..228B} with a 3-sigma clipping function\footnote{This version of SWarp is modified to have 3-sigma clipping function by S. Foucaud (private communication).} for stacking images.
The images were scaled to have a photometric zero-point of 30.000 in AB magnitudes.
Before the stacking, images were re-sampled using a Lanczos-3 6-tap filter with a 128 pixels mesh by Swarp.
A left panel of Figure \ref{fig:FWHMDistribution} shows the FWHM distribution of the individual images used for the stacking measured by the Elixir.
We can see that the FWHM ranges for $r^{'}$ -- $z^{'}$-bands are 0.60 -- 1.20, and that for $g^{'}$-band is slightly worse (0.75 -- 1.25) compared with the others. 
The FWHMs of final stacked images for $r^{'}$ -- $z^{'}$ bands are $\sim$ 0.85 -- 0.9 arcsec, and 0.96 for $g^{'}$-band.

In this work we carefully exclude a false source contamination of our final catalogue. During this data reduction, we removed peaked isolated noise using the normalized Median Absolute Deviation \citep[MAD;][]{1990AJ....100...32B}.
We considered 15 pixels surrounding each pixel in each image, and calculated a median and MAD values.
If the original pixel value deviated by more than 3 MAD from the median value, it was regarded as bad, and was replaced with the median value. 
Since fake sources lying astride over a few pixels or along artificial structures close to bright sources cannot be removed by the MAD method, we use those weight maps during source extraction by SExtractor (see $\S$\ref{sec:SExtractor}) and carry out band-merging procedure (see $\S$\ref{sec:Band Merging and Quality Frag}) to reduce the false source contaminations.

\subsection{CFHT/WIRCam (Near-Infrared: $Y$, $J$, and $K_{\rm s}$-bands)}
\label{sec:wircam}
We used the CFHT/WIRCam to obtain deep near-infrared ($Y$, $J$, and $K_{\rm s}$) images at the NEP-Deep field.
WIRCam uses four $2048 \times 2048$ HAWAII2RG CCD arrays.
The camera provides 0.3 arcsec per pixel sampling, and the close packing enables coverage of an almost contiguous field of view of $20.5^{'} \times 20.5^{'}$.
Observations in $Y$, $J$, and $K_{\rm s}$-bands were conducted in queue mode spread over 19 photometric nights from 2010 March 31 to 2010 June 29, resulting in a total of 2009 frames (PI: T. Goto).
We observed four positions with dithering, covering completely a 47$^{'}\times44^{'}$ field of view in each band.

The pre-processing, removal of the instrumental signature and astrometry for raw WIRCam images, was carried out by the $^{'}$I$^{'}$iwi\footnote{{\small http://cfht.hawaii.edu/Instruments/Imaging/WIRCam/IiwiVersion1 Doc.html}} preprocessing pipeline. 
In general, photometric calibration by SCAMP is based on both internal pairing among frames of overlapping regions with SExtracted source catalogues of WIRCam itself and external pairing with reference catalogues with SExtracted source catalogues of WIRCam. 
However, we found the internal pairing method did not work well when cross checking the uniformity of the photometric zero point across the SWarp stacked images by using the SCAMP photometric solution.
Therefore, in this paper, we only used the external pairing method.
Photometric calibration of the $J$ and $K_{\rm s}$-band observations were performed by matching to the corresponding 2MASS catalogue, the weighted images were then stacked using the modified SWarp. 
We compared 2MASS sources with photometric errors of less than 0.05 mag, $J<16.48$ mag and $K_{\rm s}<15.90$ mag, respectively, with the extracted sources of WIRCam.
We found the zero point and its uniformity within the 47$\times$44 arcmin$^2$ area for the $J$ and $K_{\rm s}$-bands to be 25.964 mag (0.072 mag) and 26.927 mag (0.073 mag), respectively.

\begin{figure}[t]
\begin{center}
  \includegraphics[width=60mm, angle=-90]{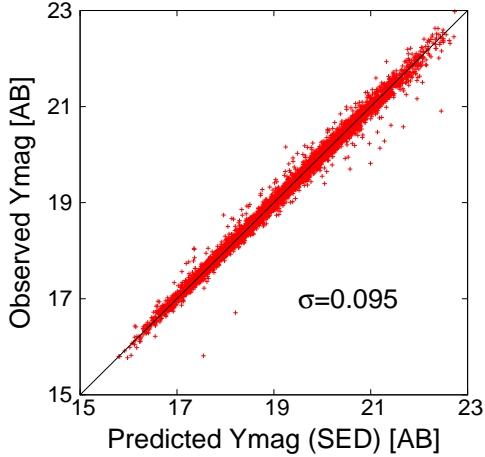}
  \caption{Comparison between the observed $Y$-band magnitude and predicted $Y$-band magnitude estimated by SED fitting of stars. From this comparison, the photometric zero point and the RMS in the $Y$-band are 25.715 mag and 0.095 mag for $Y<23$mag, respectively.}
  \label{fig:Yzeromag}
\end{center}
\end{figure}

Since the 2MASS catalogue does not have $Y$-band magnitudes, we did the photometric calibration of the $Y$-band by comparing with the predicted $Y$-band magnitudes estimated by SED fitting of stars. For the purpose of creating the predicted $Y$-band magnitude catalogue as the external reference, first we created a band-merged catalogue (see \S \ref{sec: z base catalogue}) and selected stars with a stellarity index given by SExtractor, CLASS\_STAR parameter in the $z^{'}$-band image, and appropriate $u^{*}-J$ and $g^{'}-K_{\rm s}$ colours (see \S\ref{S-G separation}).
We calculated the predicted $Y$-band magnitude based on the $\chi^2$ fitting analysis of the SED using a publicly available $LePhare$ code, \citep{2006A&A...457..841I, 2007A&A...476..137A, 2009ApJ...690.1236I} with 154 star templates, which include main sequence stellar SEDs from \cite{1998PASP..110..863P} and white dwarf SEDs from \cite{1995AJ....110.1316B}.
For the SED fitting, we input the photometric data sets from $u^{*}$ to $K_{\rm s}$-bands omitting the $Y$-band.
Then we calibrated the photometry with the predicted $Y$-band magnitude as the external reference catalogue of SCAMP, and made the final images with the modified SWarp.
In Figure \ref{fig:Yzeromag}, we compared the observed $Y$-band magnitudes from the final image with the predicted magnitudes based on the SED fitting for the selected stars.
The figure shows that both values are agreed with each other.
We obtained the photometric zero point and the root mean square (RMS) in the $Y$-band as 25.715 mag and 0.095 ($Y<23$mag).

Before the final stack, we re-samplied WIRCam images using the Lanczos-3 6-tap filter with a 128 pixels mesh by Swarp as well as MegaCam imagers. 
And then we adopted the threshold of FWHM $<$ 1.2 arcsec estimated by SExtractor to removed the bad quality images.
The FWHM distributions we used for the final stacked images are shown in the right panel of Figure \ref{fig:FWHMDistribution}, and the seeing of the final stacked images of $Y$, $J$, and $K_{\rm s}$-bands are all $\sim$ 0.7 -- 0.8 arcsec.

\section{PROCEDURE for SOURCE EXTRACTION}
\label{sec:PROCEDURE for SOURCE EXTRACTION}

\subsection{Source Extraction}
\label{sec:SExtractor}

\begin{table}[t]
\small
\begin{center}
\caption{Parameters of SExtractor.}
  \begin{tabular}{lc}
  \hline
\hline
Name & Value\\ 
\hline
DETECT\_THRESH &1.5\\
DETECT\_MINAREA & 5\\
DEBLEND\_MINCONT & 0.001\\
FILTER\_NAME    &  gauss\_3.0\_7x7.conv\\
WEIGHT\_TYPE   &   MAP\_WEIGHT\\
\hline
\hline
\end{tabular}
\label{tb:para-sex}
\end{center}
\end{table}

\begin{figure}[b]
\begin{center}
\includegraphics[width=90mm]{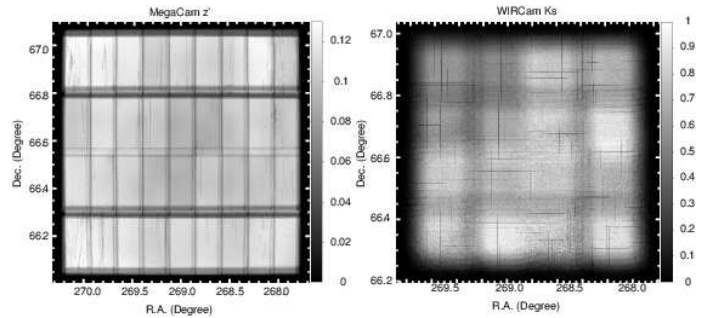}
  \caption{Examples of weight maps for MegaCam $z^{'}$-band (left) and $K_{\rm s}$-band (right), respectively. The pixel values are inversely proportional to the local variance of the image, and the areas of the pixel value of 0 are the regions without data. Crosses of clear less-weighted areas in MegaCam data are present due to the gaps between detectors, while WIRCam data are relatively homogeneous. 
 }
   \label{fig:WeightMaps}
\end{center}
  \end{figure}

\begin{table*}[t]
\small
\begin{center}
\caption{Summary of properties of the extracted source catalogues for each band.
Line (1): FWHM of PSF.
Line (2): Integration time.
Line (3): Number of detected sources in each band image. 
Line (4): Limiting magnitude in each band with a threshold of 4$\sigma$ detection within an 1 arcsec radius.
Line (5): 50\% completeness limit.
Line (6): Magnitude difference between MAG\_APER with 1 arcsec radius and MAG\_AUTO for objects at $0.45<z<0.55$.}
  \begin{tabular}{ccccccccc}
  \hline
\hline
&\multicolumn{4}{c}{MegaCam} & &\multicolumn{3}{c}{WIRCam}\\
\cline{2-5}\cline{7-9}
&$g^{'}$ & $r^{'}$  & $i^{'}$ & $z^{'}$ && $Y$ & $J$ & $K_{\rm s}$\\
\hline
FWHM (arcsec) & 0.96 & 0.88 & 0.84 & 0.86 && 0.80 & 0.81 & 0.66\\
Total integration time [ks]& 23.2&17.6&27.6&23.2&&21.2&19.9&18.3\\
Detection number  & 224,449 & 178,191 & 143,318 & 88,666 && 39.063 & 34,138  & 38,053 \\
Limiting mag [AB] & 26.7 & 25.9 & 25.1 & 24.1 && 23.4 & 23.0 &22.7\\
50\% Completeness [AB] & 26.1 & 25.4 & 24.9 & 23.7& & 22.8 & 22.4 & 22.3\\
MAG\_APER -- MAG\_AUTO [AB] & 0.55 & 0.56 & 0.49 & 0.53 & &0.48 & 0.47 & 0.38\\ 
\hline
\hline
\end{tabular}
\label{tb:photometry}
\end{center}
\end{table*}

\begin{figure*}[t]
\includegraphics[width=40mm, angle=-90]{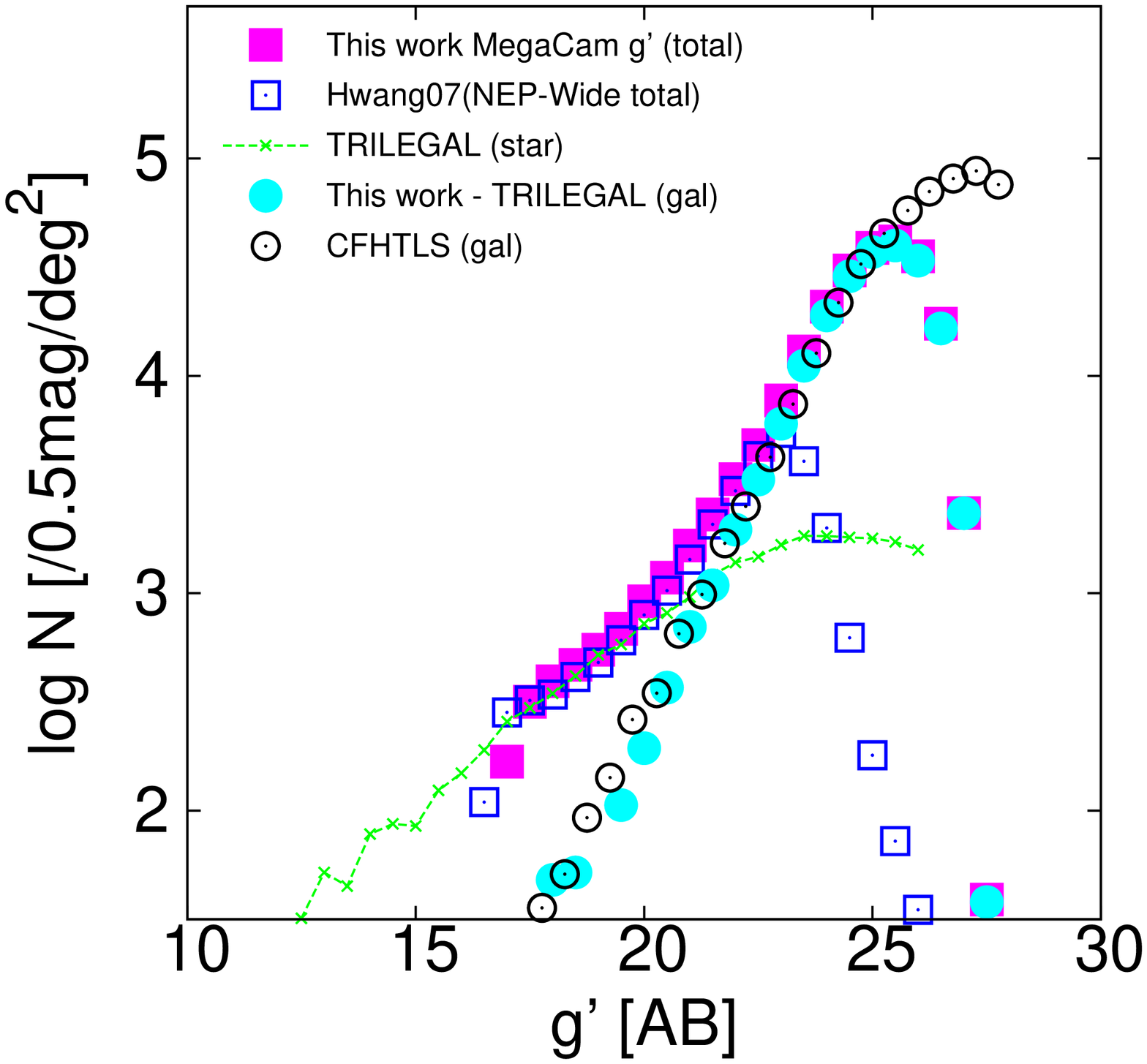}
\hspace{-1.7cm}
\includegraphics[width=40mm, angle=-90]{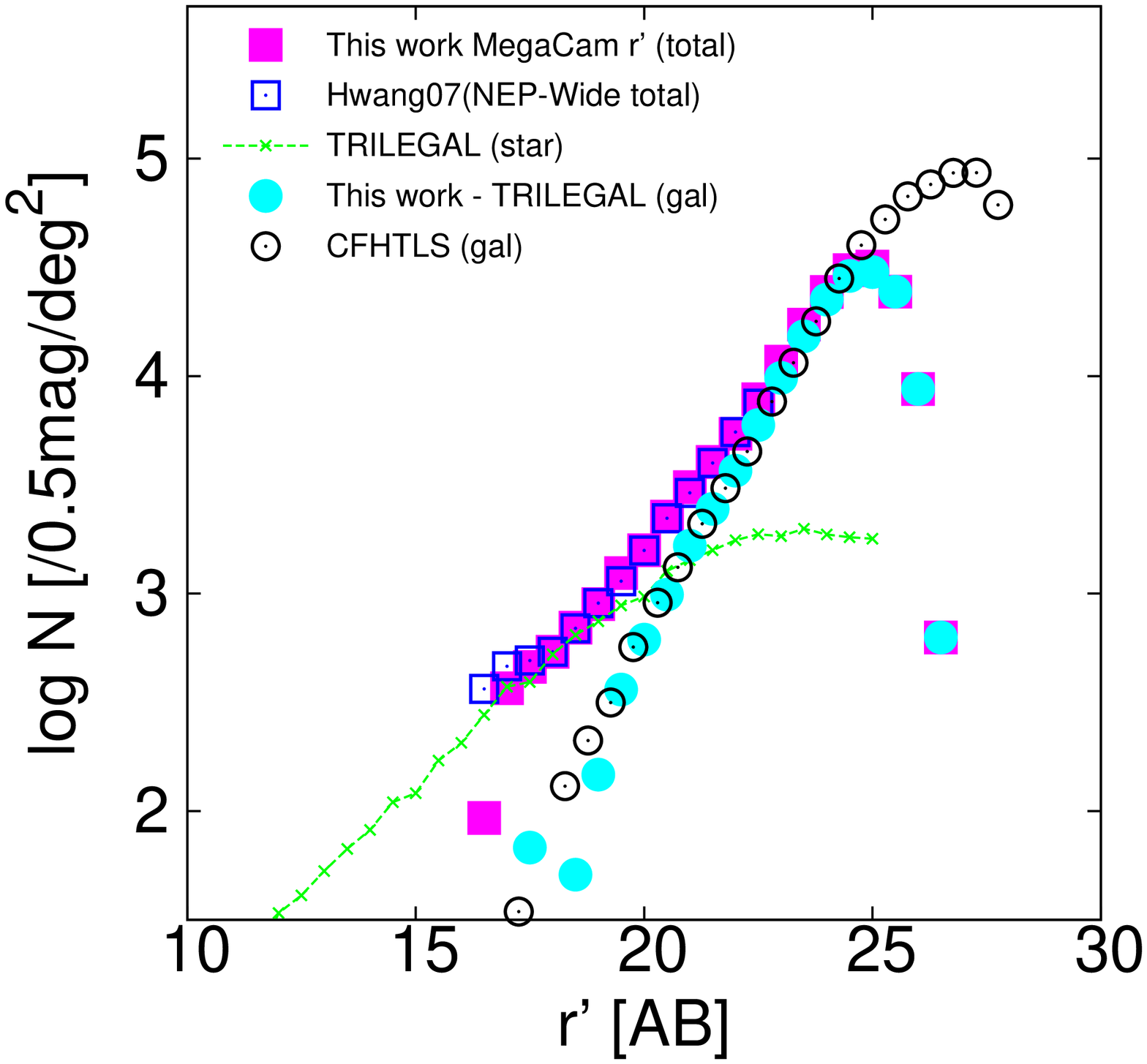}
\hspace{-1.7cm}
\includegraphics[width=40mm, angle=-90]{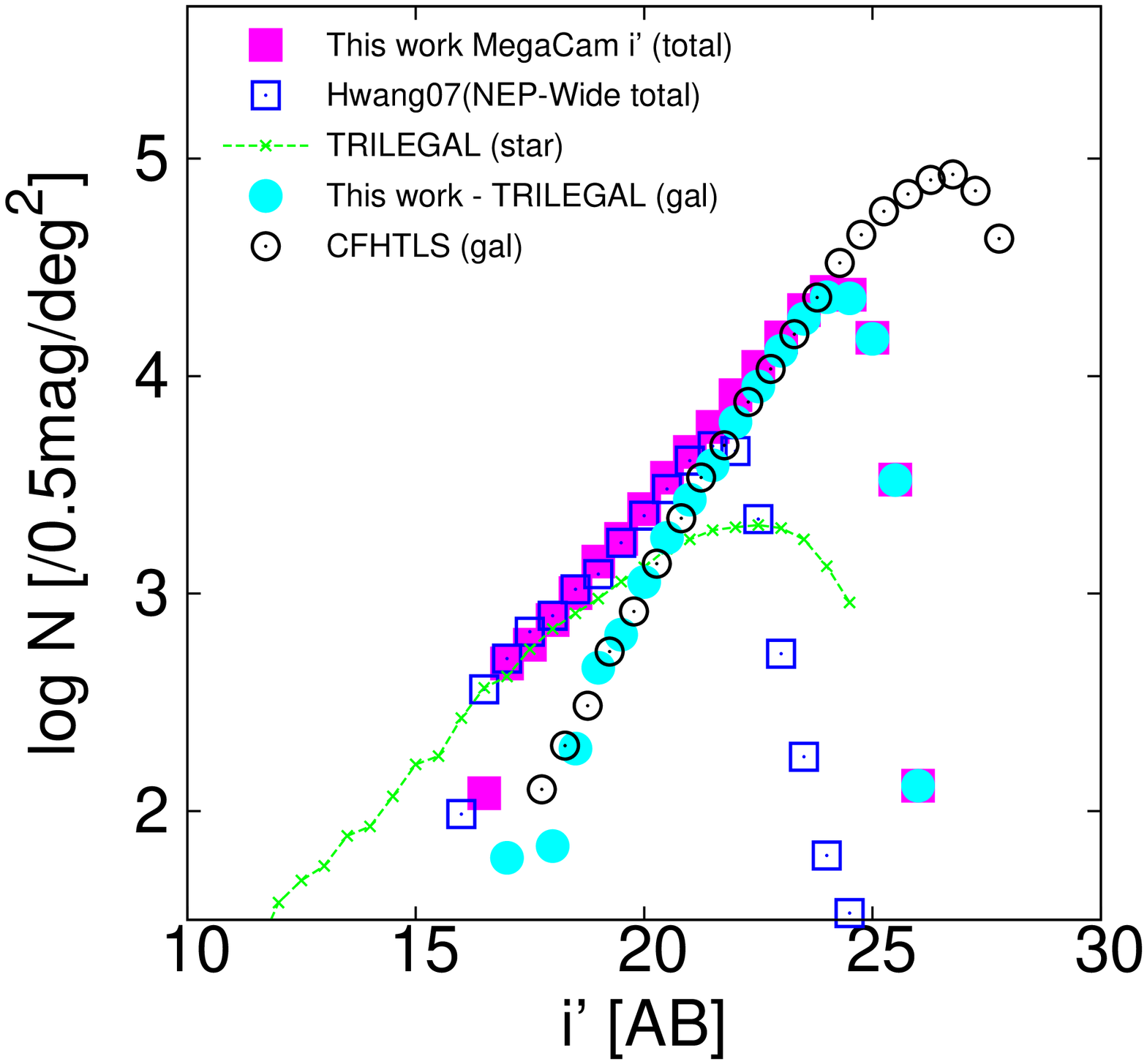}
\hspace{-1.7cm}
\includegraphics[width=40mm, angle=-90]{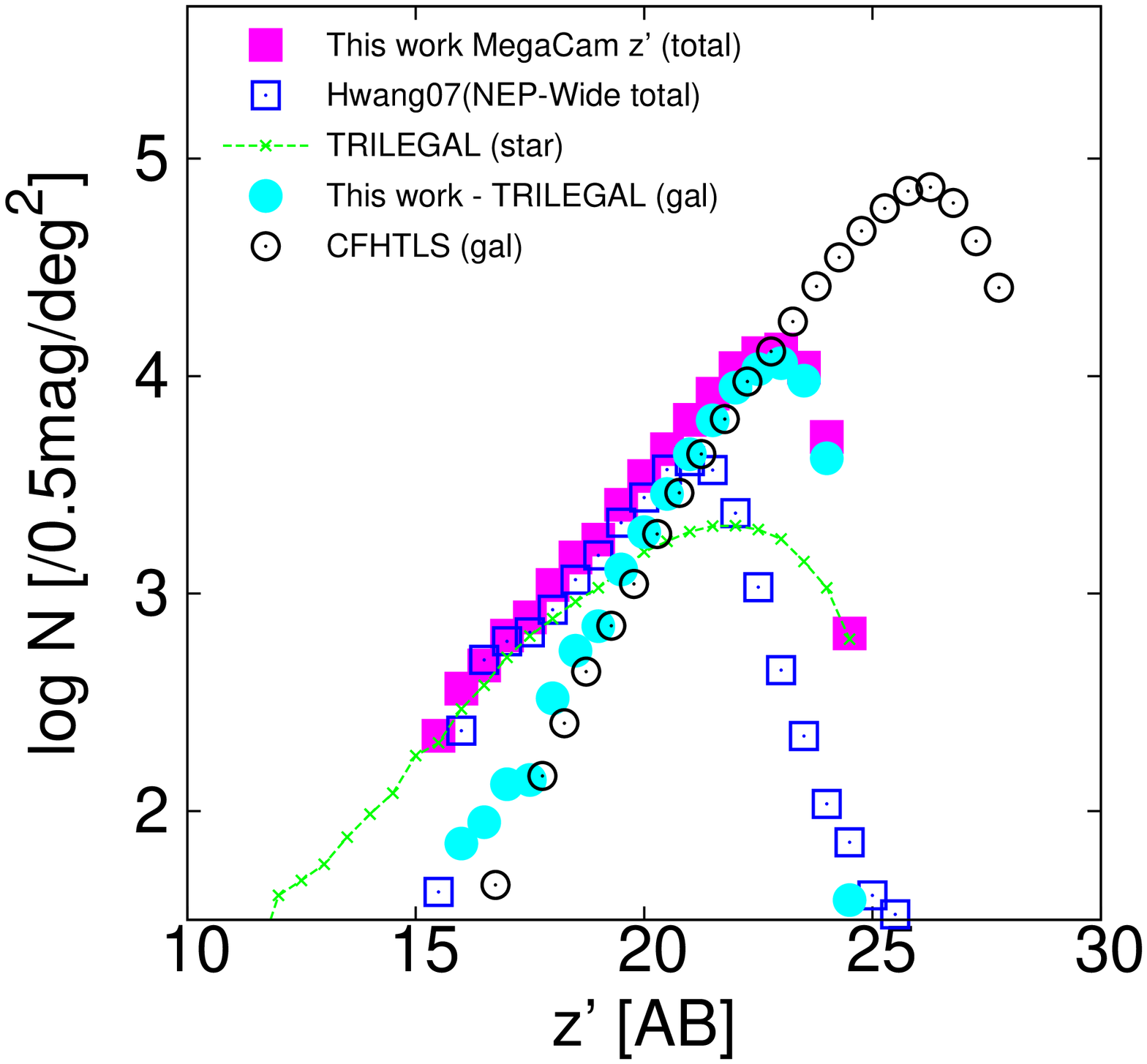}\\
\includegraphics[width=40mm, angle=-90]{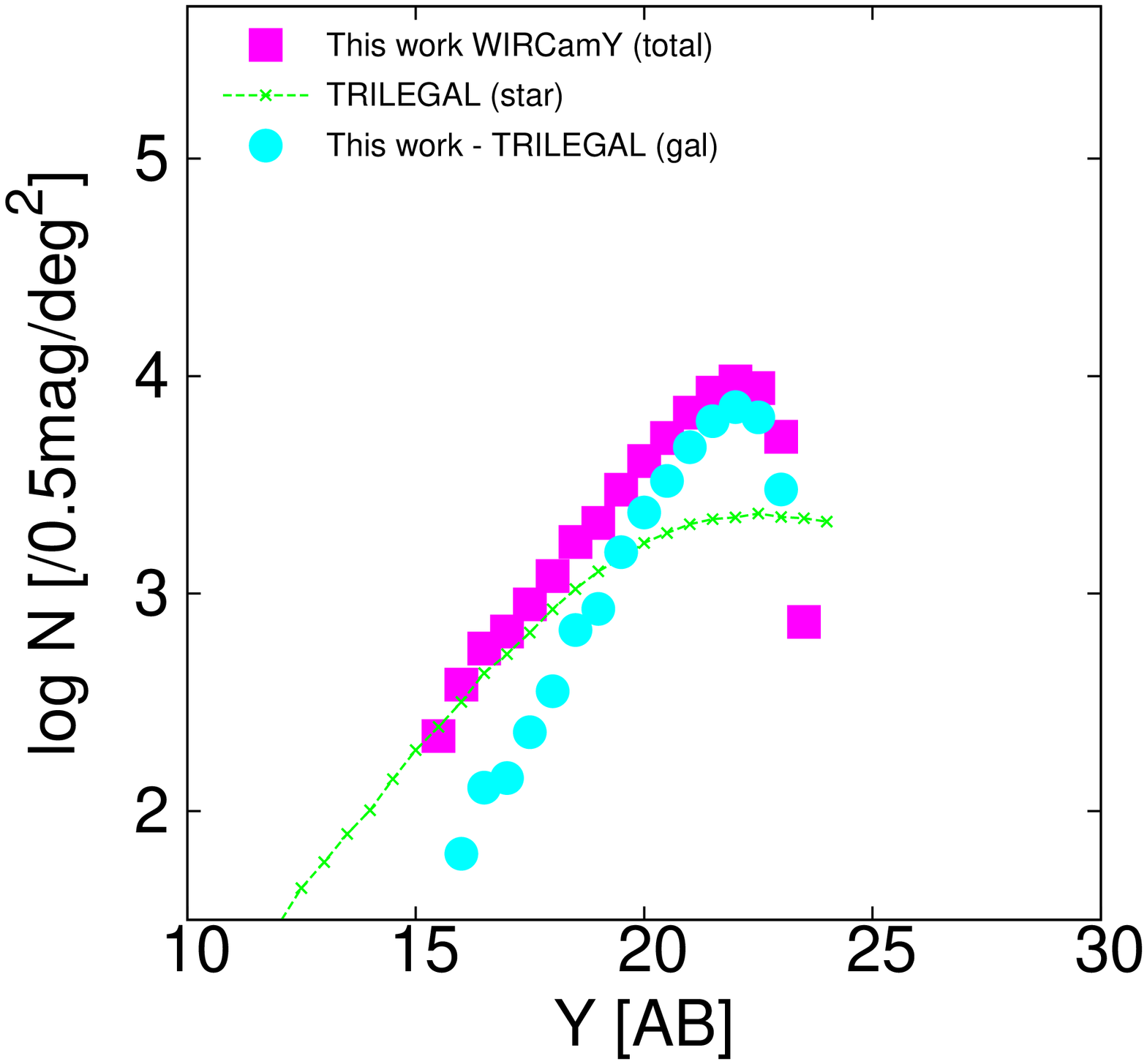}
\hspace{-1.7cm}
\includegraphics[width=40mm, angle=-90]{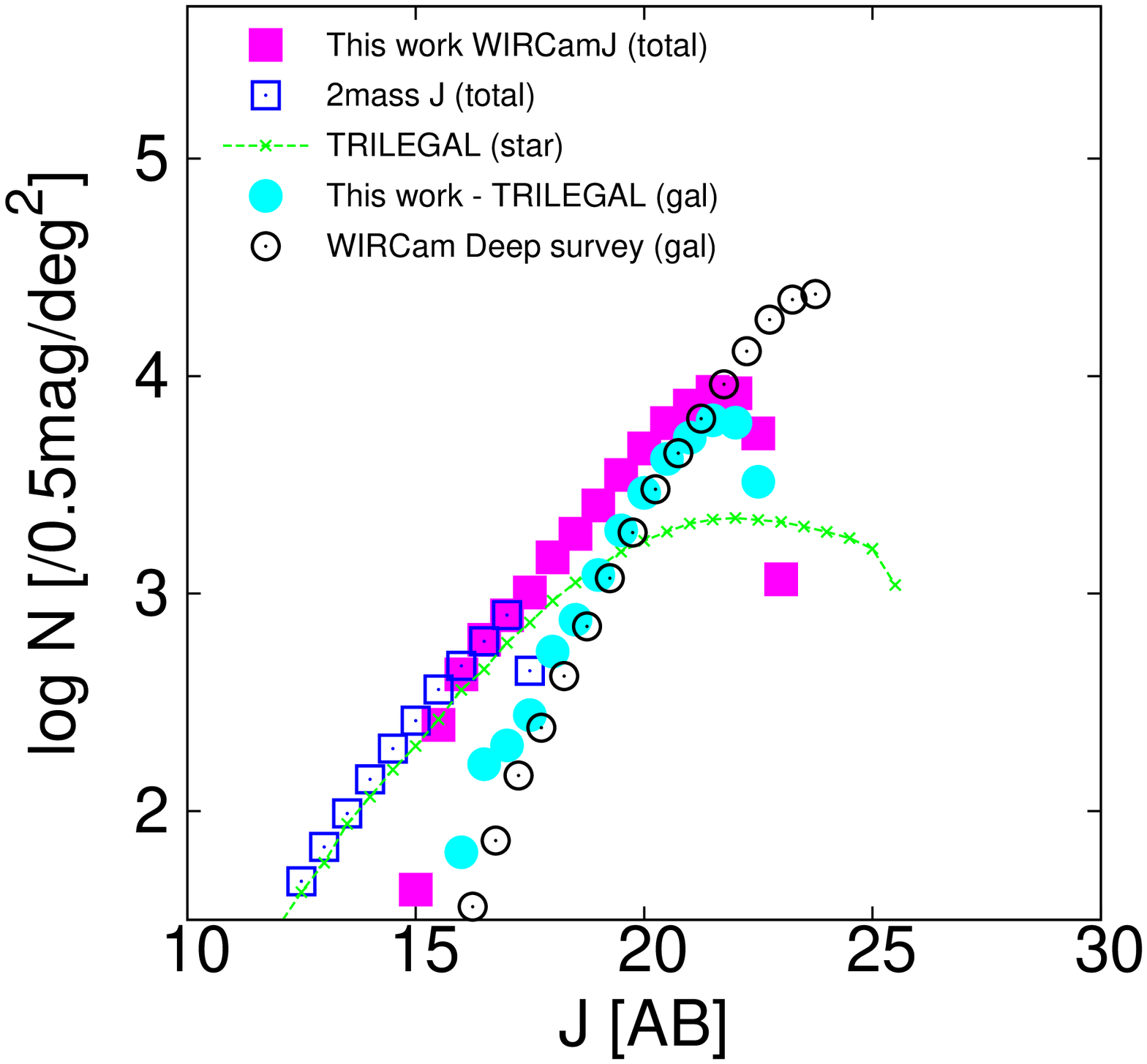}
\hspace{-1.7cm}
\includegraphics[width=40mm, angle=-90]{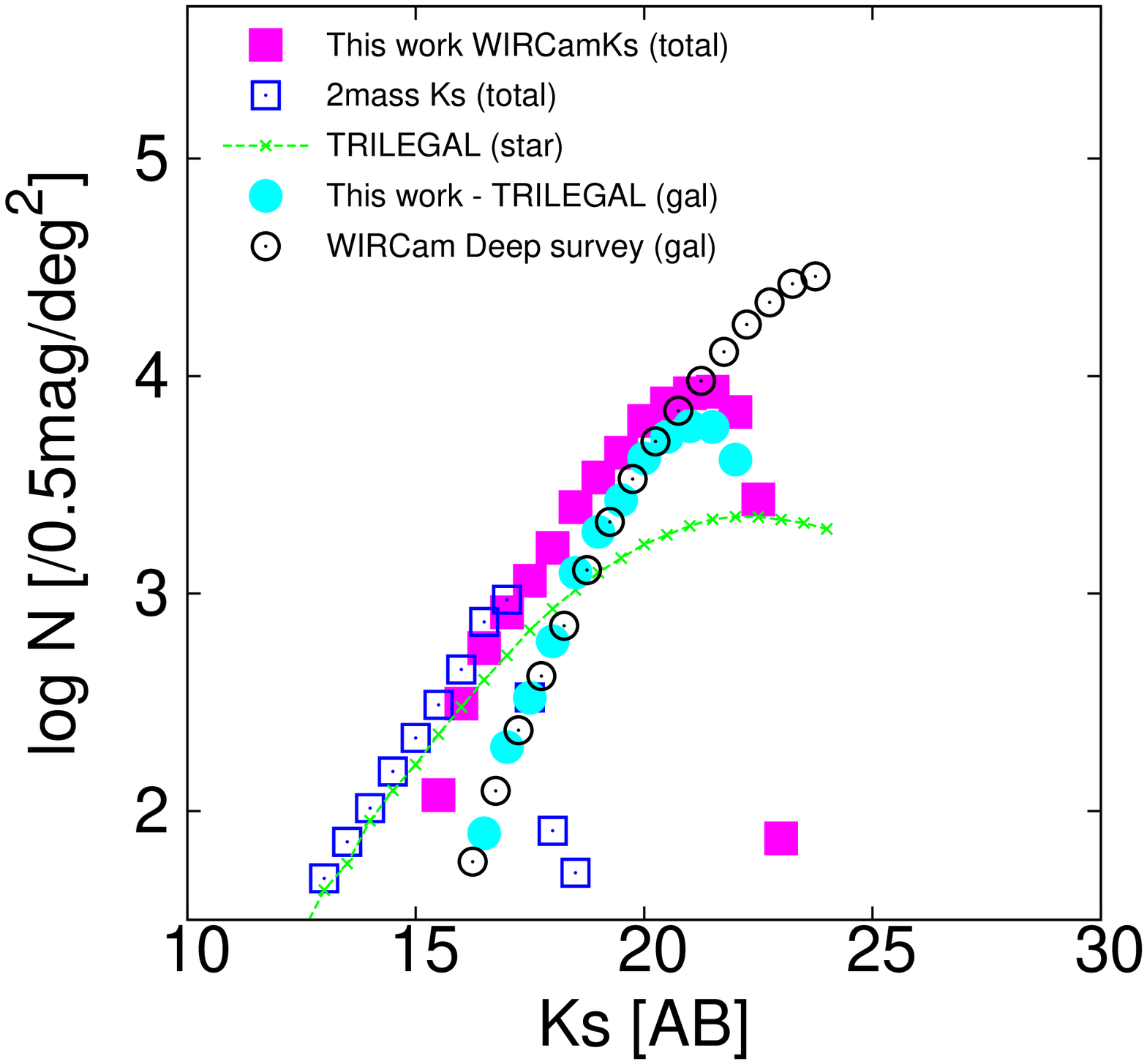}
\hspace{-1.7cm}
\includegraphics[width=40mm, angle=-90]{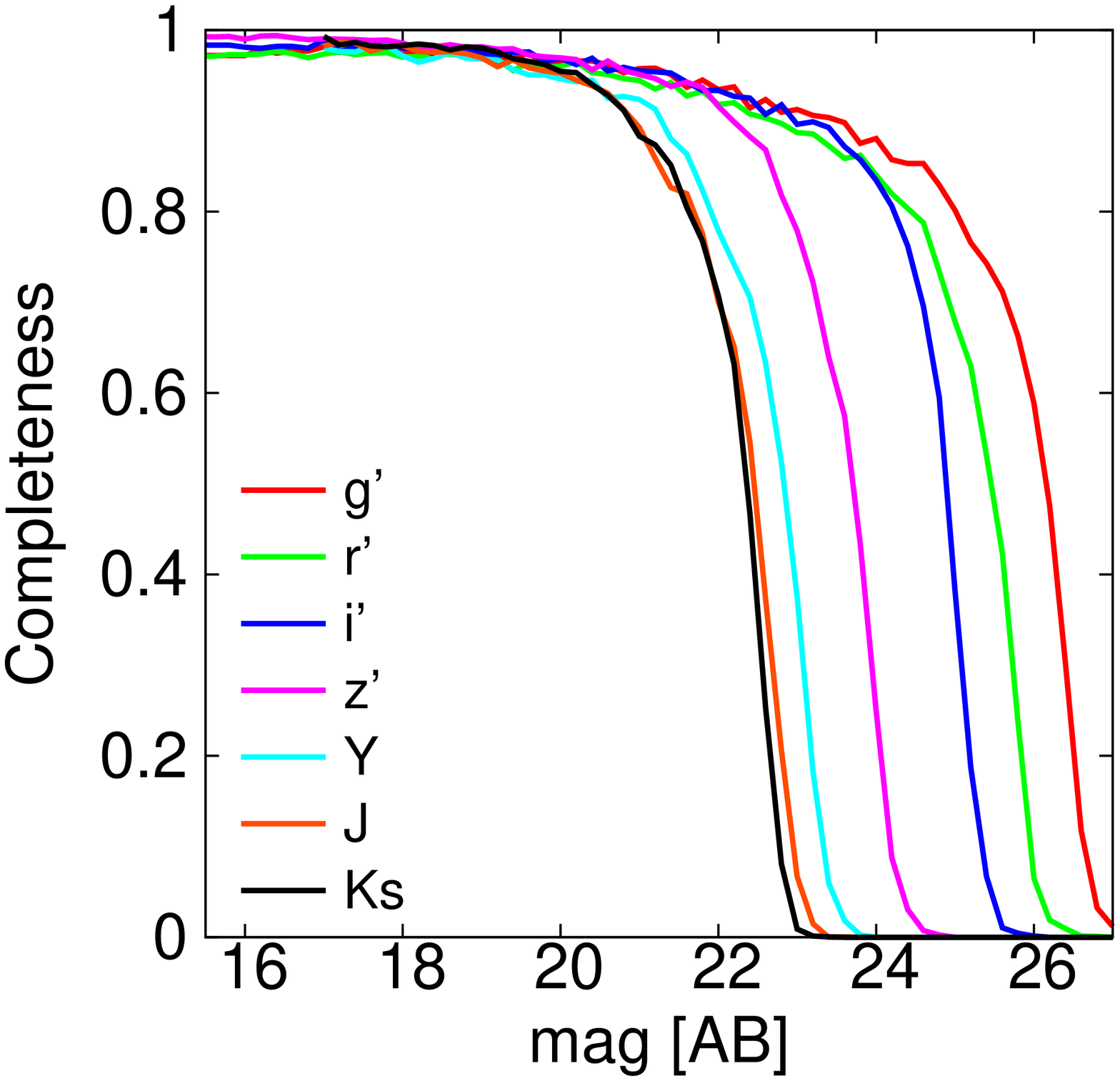}
  \caption{Number counts (all panel but right bottom one) and completeness (right bottom panel) for $g^{'}$, $r^{'}$, $i^{'}$, $z^{'}$, $Y$, $J$, and $K_{\rm s}$-bands.
In the number count panels, pink filled squares represent total number counts at NEP-Deep field (this work), and green lines show a number count of stars around NEP-Deep field using a TRILEGAL galaxy model \citep{2012rgps.book..165G}. Cyan filled circles represent number counts of galaxies which is estimated with total minus stellar number counts.
 As references, we over plotted total number counts at NEP-Wide field \citep[blue squares, ][]{2007ApJS..172..583H} and galaxy number counts from CFHTLS T0007 data for optical bands, while total number counts from 2MASS and galaxies one from WIRCam Deep survey \citep{2012A&A...545A..23B} for near-infrared. 
In the right bottom panel, red, green, blue, pink, cyan, orange, and black lines represent the completeness of $g^{'}$ -- $K_{\rm s}$, respectively.
 }
   \label{fig:NumbercountCompleteness}
  \end{figure*}

For the detection of objects in MegaCam and WIRCam images, we used SExtractor.
Seeing is not homogeneous across bands (see Figure \ref{fig:FWHMDistribution} and in Table \ref{tb:photometry}). 
In order to carry out photometry for each band with optimal aperture, we ran SExtractor with each band image individually with same parameter set.
The SExtractor parameter set for all bands is summarized in Table \ref{tb:para-sex}. 
A 7pix$\times$7pix Gaussian Point Spread Function (PSF) convolution filter with FWHM = 3.0 pixels was used for helping to detect faint sources. 
Fluxes of extracted sources were measured using SExtractor with elliptical apertures of 2.5 times of the Kron radius \citep{1980ApJS...43..305K}, i.e., SExtractor's MAG\_AUTO.
We used weight maps created during SWarp process to avoid false detections.
Figure \ref{fig:WeightMaps} is an example of weight maps (left: MegaCam $z^{'}$, right: WIRCam $K_{\rm s}$).
The MegaCam $z^{'}$-band weight map shows four less weighted regions in a cross direction and eight minor less weighted regions in longitudinal direction due to the detector gaps.
This trend for all our MegaCam data are very similar to that of $z^{'}$-band.
While the WIRCam $K_{\rm s}$-band weight map looks nicely uniform and no distinct less weighted gaps comparing with the MegaCam data.
The trend for $Y$ and $J$-ban data are almost the same as that of $K_{\rm s}$-band.
Data at the less weighted area are relatively noisier than the other area, and noise causes false detections.
Weight maps affect error estimates and reduce the number of false detections at the noisy region.

From extracted source catalogue for each band, we removed sources with a photometric flag of 4 or larger from the detected source list because these sources can be unreliable indicating that at least one pixel is saturated or close to the image boundary or both. 
The number of sources in the extracted catalogue for each band is given in Table \ref{tb:photometry}.
Figure \ref{fig:NumbercountCompleteness} (except bottom right panel) shows number counts for all our seven band data.
Number counts from \cite{2007ApJS..172..583H} at NEP-Wide field and those from 2MASS data around NEP region are over plotted in the figure as references of total (star + galaxy) number counts (blue open squares).
We found that our total number counts are consistent with those surveys' results.
We also calculated the number counts of stars at around NEP-region using a TRILEGAL galaxy model \citep{2012rgps.book..165G} in each band, and then we subtracted it from the total number counts to estimate galaxy number counts (cyan filled circles).
The number counts are compared with CFHTLS T0007\footnote{http://www.cfht.hawaii.edu/Science/CFHTLS/T0007/
CFHTLS\_T0007-TechnicalDocumentation.pdf} for MegaCam data and with WIRCam Deep survey \citep{2012A&A...545A..23B} for WIRCam $J$ and $K_{\rm s}$-bands.
At the bright magnitude range, our galaxy number counts in $g^{'}$-band is slightly larger compared with those from CFHTLS, while in $z^{'}$, and $J$-bands our results are slightly smaller than that of the comparison fields.
These differences are about 15\% and it can be explained an uncertainly of the TRILEGAL models.
At the fainter magnitude range, where the effect from star number counts is weak, our results and their comparisons are excellently agreed with each other.

\subsection{Detection Limits and Completeness}
We determined the 4$\sigma$ detection limits over a circular aperture of 1 arcsec radius. 
The detection limit changes depending on the source position on the mosaiced images since the exposure map is not uniform.
We present the median values over the entire field for all MegaCam and WIRCam bands in Table \ref{tb:photometry}. 
The limiting magnitudes are 26.7, 25.9, 25.1 and 24.1 mag for MegaCam $g^{'}$, $r^{'}$, $i^{'}$ and $z^{'}$-bands, and 23.4, 23.0, 22.7 mag for WIRCam $Y$, $J$, $K_{\rm s}$-bands, respectively. 
The completeness of each band was estimated via simulations with artificial sources.
We produced 2,000 artificial sources per 0.2 magnitude with the averaged radial profile of 20 point sources in each band. 
We separated the artificial sources by at least 60 pixels from each other. 
We ran SExtractor on each image with the artificial sources by utilizing exactly the same extraction parameters as for the original images.
A success of the detection of those artificial sources is judged based on source location (within 1 arcsec from the true location) and magnitude (within 0.5 mag from the true magnitude).
The 50\% completeness limits are listed in Table \ref{tb:photometry} and the completeness curves are shown in right panel of Figure \ref{fig:NumbercountCompleteness} as a function of magnitude.
The completeness curves show that the detection probability begins to drop rapidly at around 90\% value, and the magnitude difference between 90\% and 10\% completeness is about 2 mag.

The limiting magnitudes listed in Table \ref{tb:photometry} are actually for point sources because the values represent the 4$\sigma$ detection within an 1 arcsec aperture radius.
Over 99\% of the flux of a point source is included within the aperture (FWHM = 0.8--0.9 arcsec), while an extended galaxy at low-redshift with magnitude equal to the limiting magnitude cannot be detected with 4 $\sigma$ since such galaxies are spatially resolved in our images.
We measured the difference between the 1 arcsec radius aperture magnitude (MAG\_APER) and the total magnitude (MAG\_AUTO).
In $\S$\ref{sec:PHOTOMETRIC REDSHIFT}, we calculate the photometric redshift via SED fitting and find the peak redshift distribution of our data to be $z\sim0.5$.
We then compared the MAG\_APER with MAG\_AUTO for galaxies at $z\sim0.5$ ($0.45<z<0.55$) as a typical example in our data.
The differences in the all images is around 0.38--0.56 mag.
We, therefore, note a limiting magnitude for extended sources is typically brighter by $\sim$0.5 mag.

\section{BAND-MERGED catalogue based on $z^{'}$-BAND}
\label{sec: z base catalogue}

In this section, we describe the processes of making the band-merged catalogue.
One of our aims of this work is to construct a $z^{'}$-band based band-merged catalogue and find counterparts of $AKARI$ infrared sources \citep{2013A&A...559A.132M}.
We use optical data as a base-band for the band-merged catalogue since the optical data cover the $AKARI$ NEP-Deep field 
wider than the near-infrared ones.
Since a redder band is much more useful to identify $AKARI$ sources because bluer bands can be affected by heavy dust extinction, the $z^{'}$-band is adopted as the base-band, although it is shallower than the bluer bands.

Since $u^{*}$-band photometric data are important for the photometric redshift estimation in \S \ref{sec:PHOTOMETRIC REDSHIFT}, in this work we merged not only the newly obtained MegaCam 4 bands and WIRCam 3 bands, but also the $u^{*}$-band catalogue constructed by \cite{2012A&A...537A..24T}.
This $u^{*}$-band image is centered at R.A. = 17$^h$55$^m$24$^s$, Dec. = $+$66$^{\circ}$37$^{'}$32$^{"}$ [$J$2000] and has a limiting magnitude of 24.6 mag [5$\sigma$ ; AB]. 
Details of the source extraction and the photometry for $u^{*}$-band are described in \S 2.2.2 of \cite{2012A&A...537A..24T}.

\subsection{Astrometric Accuracy}
\label{sec:Astrometry}
\begin{figure}[t]
\begin{center}
  \includegraphics[width=35mm, angle=-90]{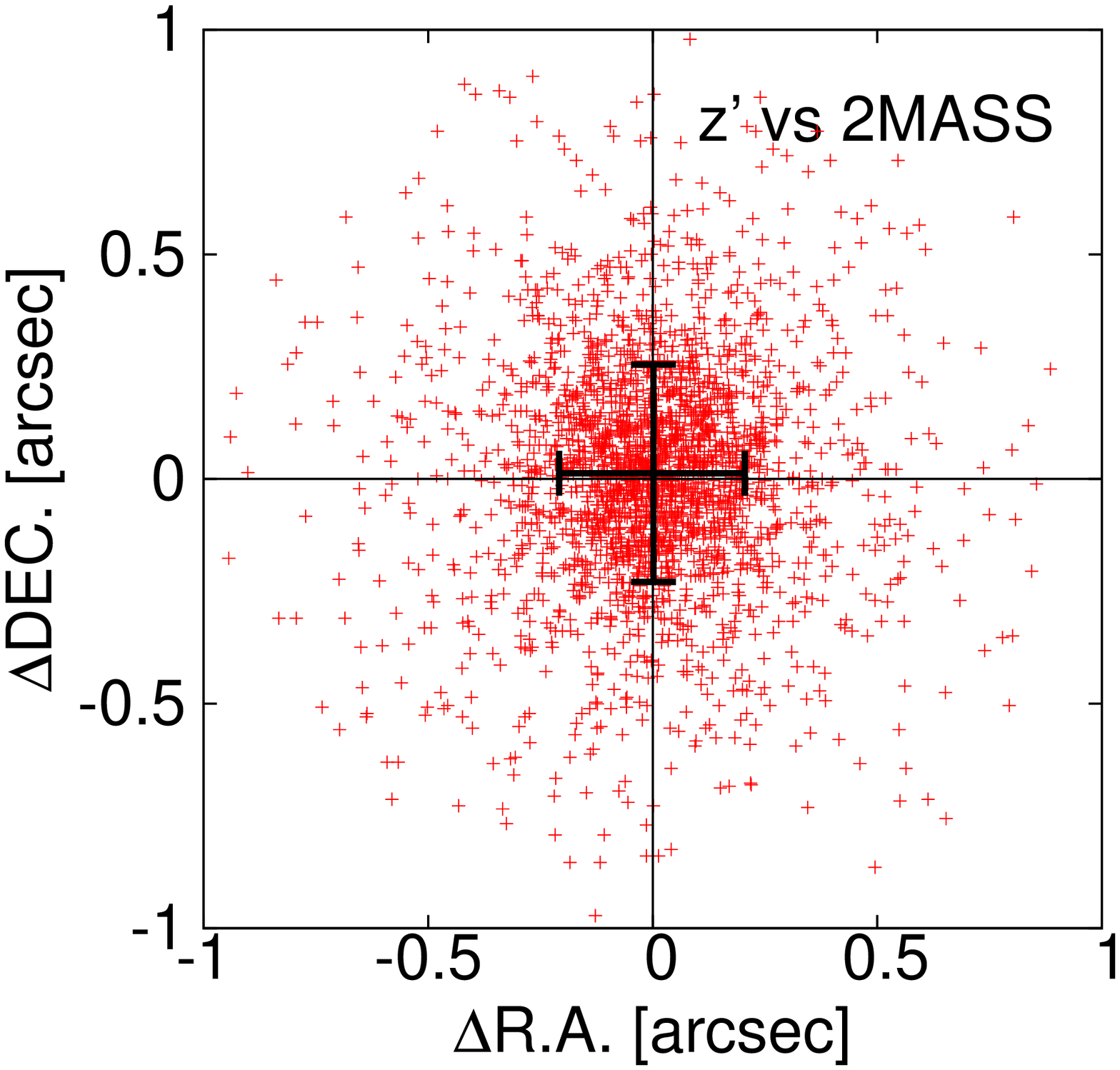}
\hspace{-1.2cm}
\includegraphics[width=35mm, angle=-90]{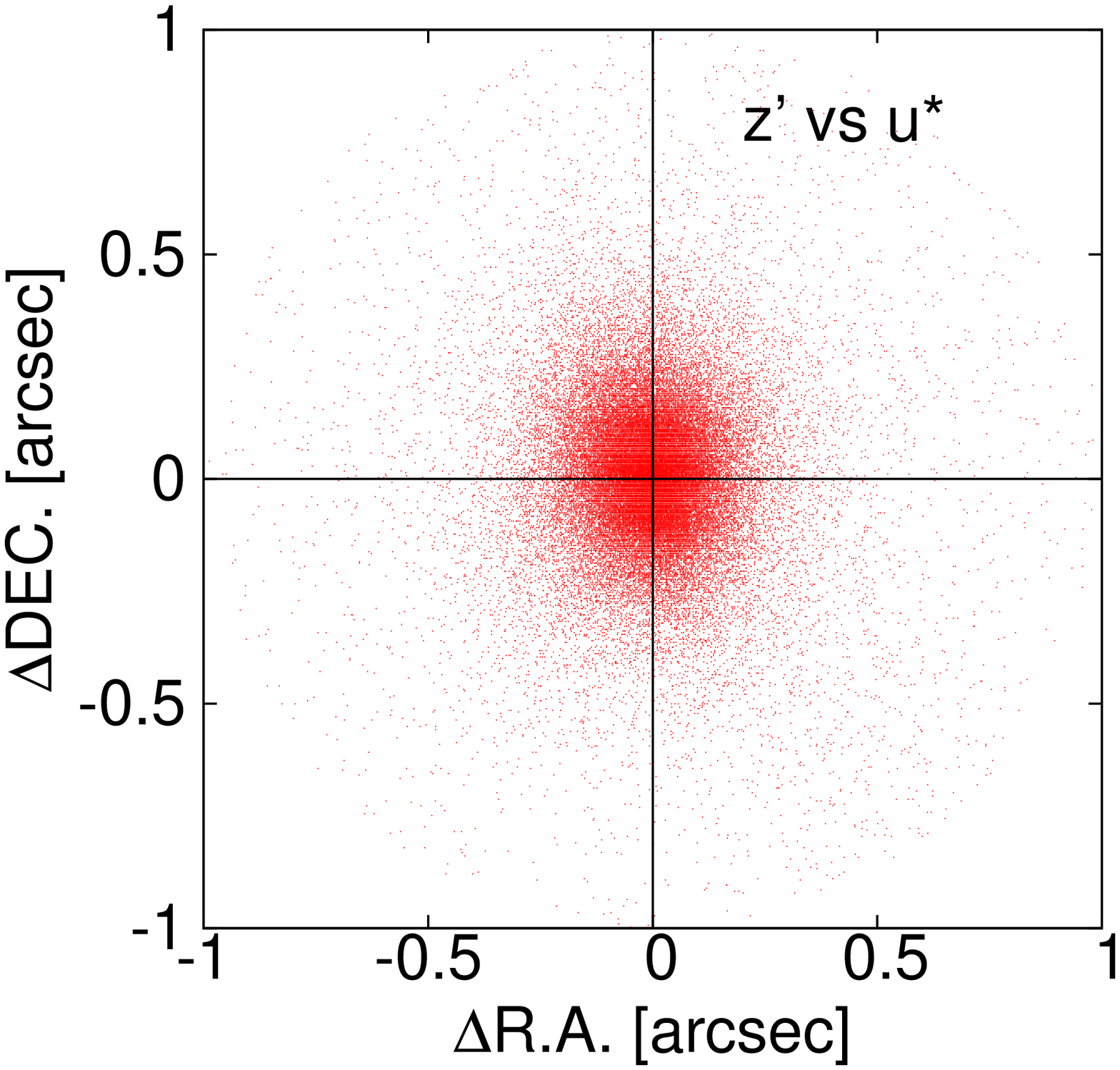}\\
  \includegraphics[width=35mm, angle=-90]{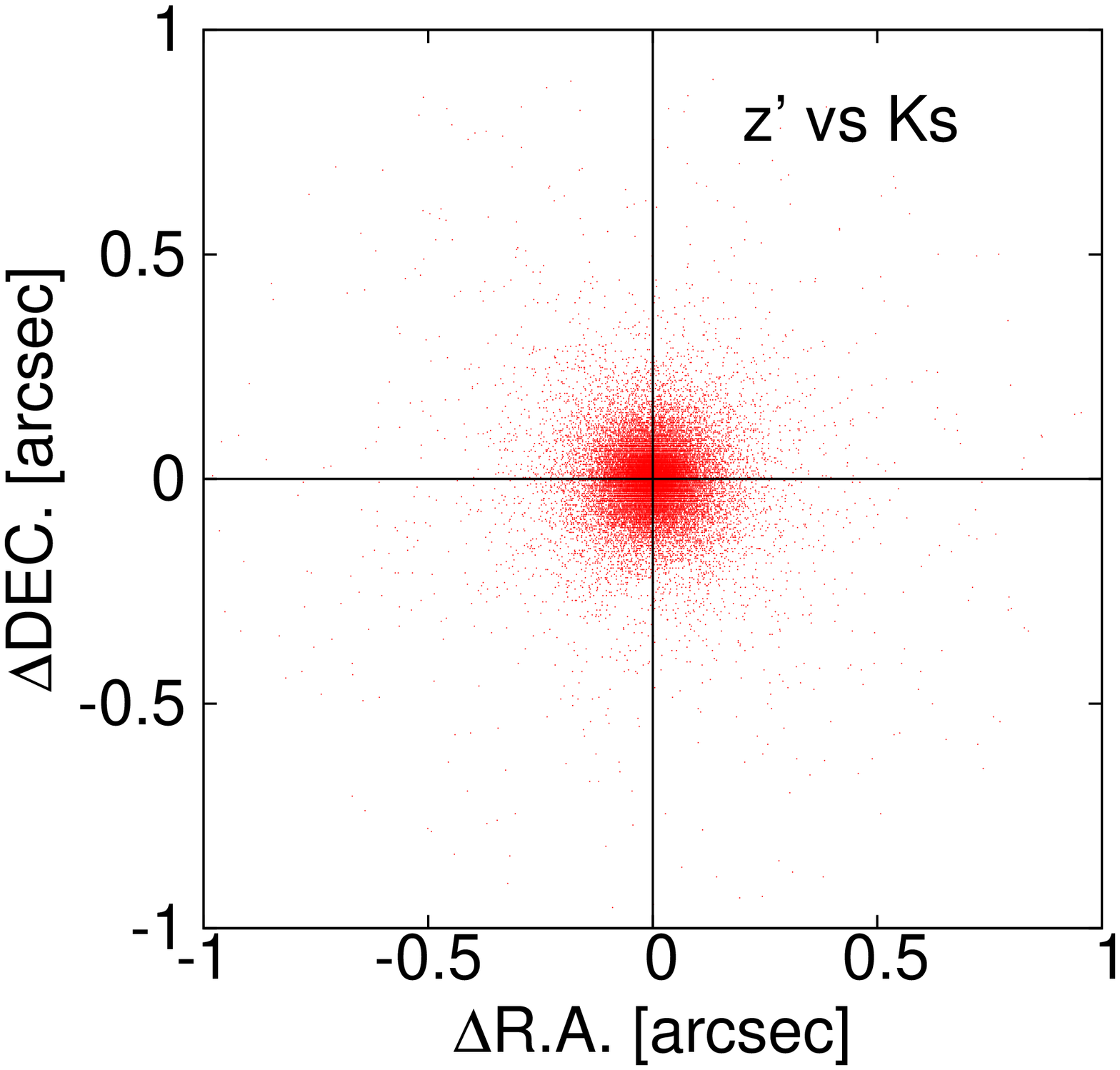}
  \caption{Relative positions of the counterparts between the $z^{'}$-band and 2MASS, $u^{*}$, and $K_{\rm s}$-band catalogues. The error bars in the top left figure indicate the root mean square for each R.A. and DEC. direction.}
  \label{fig:RADEC-Z-GRIu}
\end{center}
  \end{figure}

We cross-matched sources in all the bands to create the band-merged catalogue based on the $z^{'}$-band and checked the astrometric accuracy by examining the coordinate offset with respect to 2MASS and our other band catalogues.
Counterparts in each catalogue for $z^{'}$-band sources were searched within an 1 arcsec radius.
We found 2334 counterpart of 2MASS sources.
The spatial distribution of the offset, mean offsets and RMS offset, are 0.013 arcsec and 0.32 arcsec, respectively. 
For the other bands, the RMS for $u^{*}$-band is 0.18 arcsec, and 0.08--0.11 arcsec in the other bands.
The large RMS in $u^{*}$-band is probably due to a slightly degraded PSF.
All the offsets and the RMS offsets are summarized in Table \ref{tb:3sigma-RADECoffset}, and Figure \ref{fig:RADEC-Z-GRIu} shows examples of the relative positions between the $z^{'}$-band and 2MASS, $u^{*}$, and $K_{\rm s}$-bands.

\begin{table}[t]
\begin{center}
\caption{Accuracy of astrometry in arcsec. Column(1): Pair of bands. Column(2) and (3): R.A. offset and RMS, respectively. Column(4) and (5): DEC. offset and RMS, respectively.}
  \begin{tabular}{crccrc}
  \hline
\hline
Band &\multicolumn{2}{c}{$\Delta$ R.A.} && \multicolumn{2}{c}{$\Delta$ DEC.} \\
\cline{2-3}\cline{5-6}
&\multicolumn{1}{c}{offset}  & \multicolumn{1}{c}{RMS} && \multicolumn{1}{c}{offset}  & \multicolumn{1}{c}{RMS}\\
\hline
$z^{'}$ vs. 2MASS   &0.0018  & 0.2067 && 0.0129 &0.2420\\ 
$u^{*}$ vs. $z^{'}$  & $-$0.0043 & 0.1114 &&      0.0098 & 0.1438\\
$g^{'}$ vs. $z^{'}$   & $-$0.0041 & 0.0733 &&      0.0107 & 0.0821 \\
$r^{'}$  vs. $z^{'}$   & $-$0.0024 & 0.0645 &&      0.0032 & 0.0641 \\
$i^{'}$  vs. $z^{'}$   & $-$0.0019 & 0.0591 &&      0.0016 & 0.0583 \\
$Y$ vs. $z^{'}$        & $-$0.000003 & 0.0613 &&  0.0002 & 0.0630 \\
$J$  vs. $z^{'}$        &       0.0002 & 0.0620 &&      0.0003 & 0.0626 \\
$K_{\rm s}$  vs. $z^{'}$ & 0.0003  & 0.0623&& $-$0.0001 & 0.0628 \\
\hline
\end{tabular}
\label {tb:3sigma-RADECoffset}
\end{center}
\end{table}

\subsection{Band Merging}
\label{sec:Band Merging and Quality Frag}
To identify counterparts with cross-matching between $z^{'}$-band and the other bands, we adopted a 0.5 arcsec search radius that corresponds to 3 times the RMS of astrometric accuracy between $z^{'}$-band and $u^{*}$-band, i.e., the pair that gave the worst coordinate matching accuracy.
This small search area size helped us to minimize the chance coincidence rate (see $\S$\ref{sec: false detection}).

After the source matching, we compiled all the catalogues into a single band-merged catalogue based on the $z^{'}$-band sources. 
To ensure a robust catalogue against false sources and transients (e.g., time variable objects and asteroids), we excluded sources with only $z^{'}$-band detection.
As a result, the catalogue includes 85$\hspace{0.1em}$797 out of 88$\hspace{0.1em}$666 $z^{'}$-band sources. 
The number of matched sources between $z^{'}$-band and the other bands are summarized in Table \ref{tb:matching}.
In Table \ref{tb:final-catalogue}, we summarize the statistics of sources with multiple band detections for all 8 bands in the optical and near-infrared together. 

\begin{table*}[htpb]
\small
\begin{center}
\caption{Summery of counterpart matching. 
Line (1) represents the total number of $z^{'}$-band sources with counterparts in other bands.
The number of $z^{'}$-band sources classified into Case A, B, and C, are summarized in lines (2)--(4), respectively, for each band.}
  \begin{tabular}{ccccccccc}
  \hline
\hline
&\multicolumn{4}{c}{MegaCam} & &\multicolumn{3}{c}{WIRCam}\\
\cline{2-5}\cline{7-9}
&$u^{*}$&$g^{'}$ & $r^{'}$  & $i^{'}$ && $Y$ & $J$ & $K_{\rm s}$\\
\hline
Total                      & 55,145 & 75,961 & 79,454& 83,295 && 34,162 & 31,002 & 31,634\\
Case A (flag=0)  & 55,063 & 75,959  & 79,452 & 83,286 && 34,160 & 31,000 & 31,634 \\
Case B (flag=1) &       82   &    0         &     2      &       5    &&   0        &   0          &          0\\
Case C (flag=2) &         0   &      2       &     0      &       4    &&   2        &   2          &          0\\
\hline
\end{tabular}
\label{tb:matching}
\end{center}
\end{table*}

\newlength{\myheight}
\setlength{\myheight}{0.5cm}
\begin{table*}[htbp]
\small
\begin{center}
\caption{Number of sources with multiple band detections.}
  \begin{tabular}{ccccccccc}
  \hline
\hline
\# of detected bands&2&3 & 4  & 5 &6& 7& 8 &total\\
\hline
\# of sources & 2,583 & 3,725 & 17,214 & 28,727 & 6,108 & 8,550 & 18,890 & 85,797\\
fraction & 0.03 & 0.04 & 0.20 & 0.34 & 0.07 & 0.10 & 0.22 & 1\\
\hline
\end{tabular}
\label{tb:final-catalogue}
\end{center}
\end{table*}

During the matching of the sources, we identified three types of counterpart identifications:
(case A) a single $z^{'}$-band source has a unique counterpart, 
(case B) a single $z^{'}$-band source has multiple counterpart candidates, and
(case C) multiple $z^{'}$-band sources have one common counterpart candidate.
Due to the small searching area (0.5 arcsec radius), a significant fraction of the counterparts of $z^{'}$-band sources are categorized into the case A.
We assigned these unique pairs the $^{"}$0$^{"}$ flag to indicate that the pair is reliable in our band-merged catalogue. 
Both of the cases B and C are very rare for all bands ($<$ 0.01\%) except the case B of $u^{*}$-band. 
We visually checked the $u^{*}$-band image and suspect that distorted PSF in $u^{*}$-band causes such multiple detections of a single source. 
We treated the case B by summing fluxes of the multiple counterpart candidates and give them a flag $^{"}$1$^{"}$, while for the case C we listed the common source in other band as a counterpart for both of the $z^{'}$-band sources with flag $^{"}$2$^{"}$ in the catalogue. 
The number of pairs found in those case are summarized in Table \ref{tb:matching}.

\subsection{False Matching Rate}
\label{sec: false detection}
Reducing the risk of contamination by false sources is one of important points for catalogue construction.
For this purpose, we cleaned up bad pixels ($\S$\ref{sec:OBSERVATION and DATA REDUCTION}), used weight maps during extracting sources in each band image individually ($\S$\ref{sec:SExtractor}), and cross-matched sources among bands and excluded sources detected only in $z^{'}$-band ($\S$\ref{sec:Band Merging and Quality Frag}).
However, there is a small, but non-zero possibility of false matching in the merged catalogue.
In this section we estimate the false matching rate in the catalogue.

First, we estimated the false detection rate in each image.
We multiplied each image by --1 to make a positive-negative reversed image, and then ran SExtractor on a negative image with the same parameters as for a positive image to detect false sources.
The main locations of the false sources in the negative image are along the radial structures of bright stars and at edges of the image.
The numbers of detections in the negative image for each band are summarized in Table \ref{tb:negative-matching}.
In the $z^{'}$-band case, we detected 88$\hspace{0.1em}$666 and 943 sources in the positive and negative images, respectively.
Thus the false detection rate for $z^{'}$-band is $\sim$1.06\%.
In our worst case, $r^{'}$-band, the false detection rate is $\sim$1.65\%.

Second, we estimated a false--false matching rate between $z^{'}$ and other band spurious sources in the negative images with 0.5 arcsec search radius.
Even in the worst case, between $z^{'}$ and $g^{'}$-bands, the number of spuriously matched sources is only 15 (15/943$\sim$1.60\%). This small percentage is probably due to the fairly tight matching radius criterion.

Next, we measured a real--false matching rate between the sources in the positive $z{'}$-band image and false sources in other band negative images.
Even in the worst case, $z^{'}$ and $u^{*}$-bands, the rate is very low ($\sim$0.002\%).
The results of false matching are also shown in Table \ref{tb:negative-matching}. 
Surprisingly, the real--false matching rate is smaller than the false--false matching rate in all bands except the $u^{*}$-band.
This is because most false sources detected in the negative images locate around strong artifacts around bright stars and the sky noise itself in each image randomly created less spurious sources.
These results suggest that the real--false matching rate is extremely small in the final catalogue when we use the 0.5 arcsec radius search area, nevertheless many of the faint sources appear around the artifacts and the edges of images.

Finally, we calculated the probability of chance coincidences between the real sources.
If their source positions are completely uncorrelated and randomly distributed, the chance coincidence rate can be calculated with $n\times S$, where $n$ is number density of images and $S$ is search area.
In the case of the $g^{'}$-band, $n$ is 224$\hspace{0.1em}$450 per square degree and $S$ is $\pi\times(0.5$ arcsec)$^{2}$, giving a predicted chance coincidence of 1.36\%.
However these positions are actually correlated when the sources are detected in the same field of view, 
because a source at a given band should be present in the other bands.
Therefore the real chance coincidence rate in the final catalogue is expected to be much smaller than the predictions. 
We thus conclude that the false matching probability is negligibly small.

\begin{table*}[htpb]
\begin{center}
\caption{Summary of chance coincidence rate in the final catalogue. Col(1): Band name. Col(2): The number of detected sources in the original images shown in Table \ref{tb:photometry}. Col(3): The number (rate) of the detected sources in the negative image. Col(4): The number (rate) of matching between false sources in the negative images of the $z^{'}$-band and other bands within a 0.5 arcsec radius circle. Col(5): The number (rate) of matching between the real sources in the original $z^{'}$-band image and the false sources in the negative images of other bands within a 0.5 arcsec radius. Col(6): Predicted probability of chance coincidence rate within a 0.5 arcsec radial area, assuming a spacial distribution of sources in the $z^{'}$-band and other bands are completely independent.}
 \begin{tabular}{crrrrr}
  \hline
\hline
 \multicolumn{1}{c}{BAND} & \multicolumn{1}{c}{positive} & \multicolumn{1}{c}{false} &\multicolumn{1}{c}{\# of false-false match} &  \multicolumn{1}{c}{\# of real-false match} &\multicolumn{1}{c}{predicted}\\
\hline
$u^{*}$   & 149,168  &  1,475 (0.99\%)  &    0 (0.00\%)  & 2 (0.002\%) & 0.90\%\\
$g^{'}$    &  224,449  &  2,361 (1.05\%) & 15 (1.60\%)  & 1 (0.001\%)  & 1.36\%\\
$r^{'}$     &  178,191  &  2,944 (1.65\%) & 13 (1.38\%)  & 1 (0.001\%)  & 1.08\%\\
$i^{'}$     &  143,318  &     911 (0.64\%) &   9 (0.95\%)  & 0 (0.00\%)    & 0.87\%\\
$z^{'}$    &    88,666  &     943 (1.06\%) &              N/A   &   N/A              &  N/A    \\
$Y$         &     39,063  &    456 (1.17\%) &   0 (0.00\%)   & 0 (0.00\%)   & 0.41\%\\
$J$          &     34,138  &    183 (0.54\%) &   0 (0.00\%)   & 0 (0.00\%)   & 0.36\%\\
$K_{\rm s}$     &     38,053  &     98 (0.26\%)  &   0 (0.00\%)   & 0 (0.00\%)    & 0.40\%\\
\hline
\hline
\end{tabular}
\label{tb:negative-matching}
\end{center}
\end{table*}

\section{STAR -- GALAXY SEPARATION}
\label{S-G separation}

\begin{figure}[htbp]
\begin{center}
  \includegraphics[width=35mm, angle=-90]{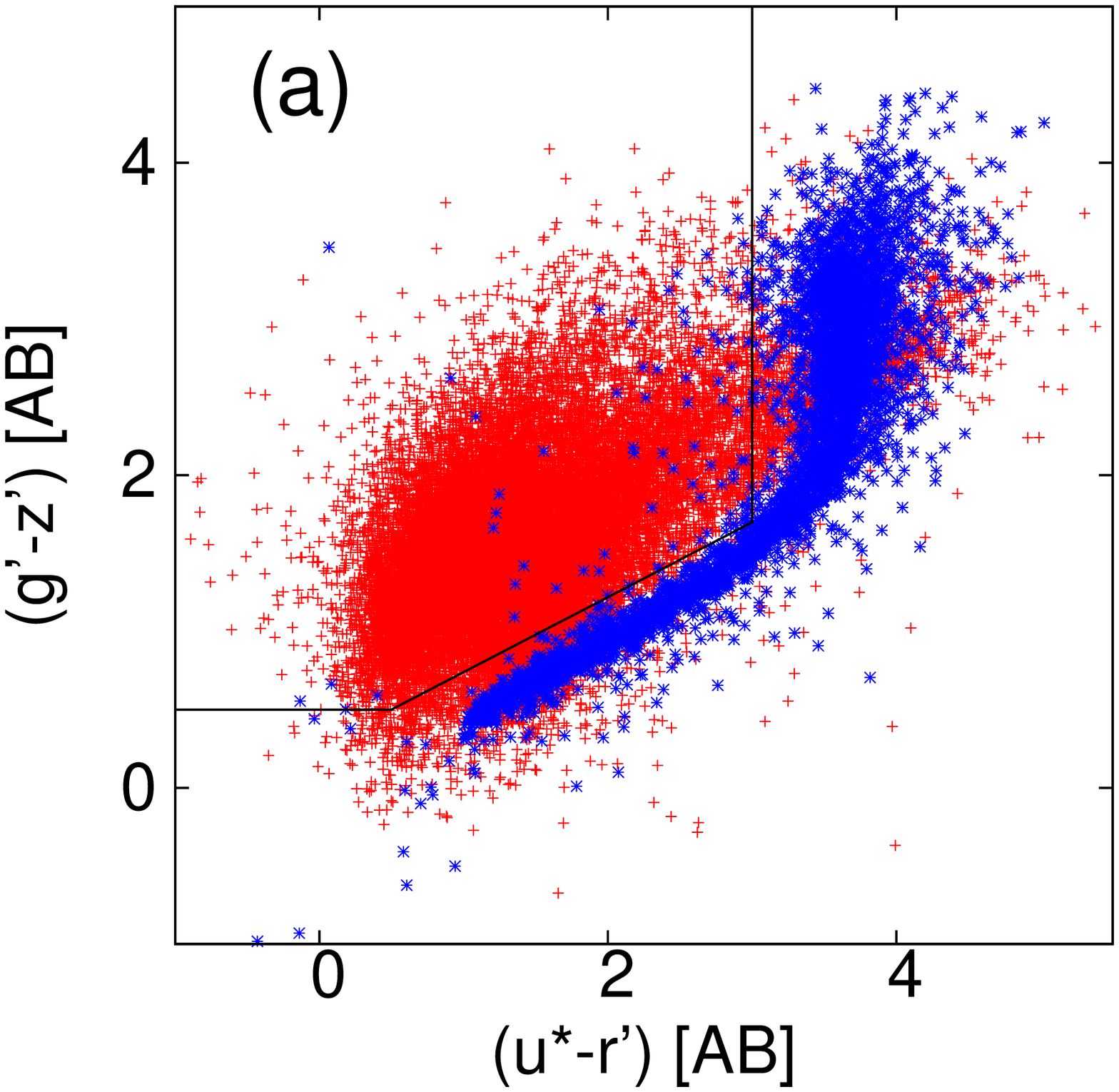}
\hspace{-1.2cm}
  \includegraphics[width=35mm, angle=-90]{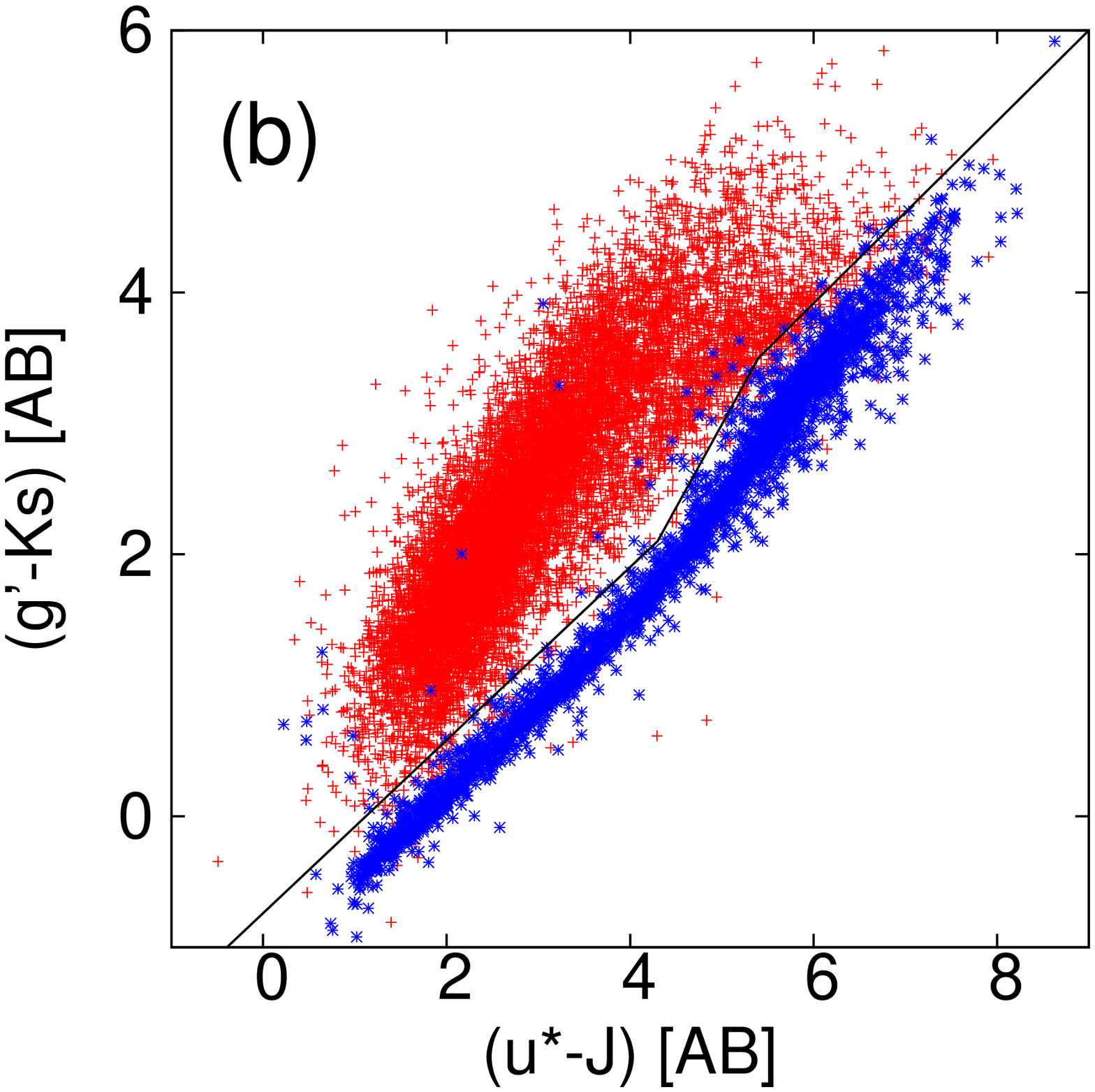}
  \caption{(Left) $u^{*}-r^{'}$ versus $g^{'}-z^{'}$ diagram. Objects classified as stars by SExtractor (CLASS\_STAR $>$ 0.95) are plotted in blue and objects classified as galaxies (CLASS\_STAR $<$ 0.2) are plotted in red. The solid line is the boundary separating the point sources from extended sources that \cite{2007ApJS..172..583H} found. (Right) $u^{*}-J$ versus $g^{'}-K_{\rm s}$ colour diagram. The same colours are used. The solid line is the boundary between point sources and extended sources we defined (see text). }
  \label{fig:urgz-uJgKs-stellar-09502}
\end{center}
  \end{figure}

In this section, we attempt to separate stars securely from galaxies using stellarity index of SExtractor (CLASS\_STAR) and colour.
CLASS\_STAR results from a supervised neural network that is trained to perform a star/galaxy classification with values between 0 and 1.
Sources with CLASS\_STAR $>$ 0.95 (as stars) and those with the value $<$ 0.2 (as galaxies) in the catalogue are plotted separately in the $u^{*}-r^{'}$ versus $g^{'}-z^{'}$ colour plane (left panel of Figure \ref{fig:urgz-uJgKs-stellar-09502}). 
The figure shows that they are clearly separated with some small fraction of exceptions by a three-straight-line boundary connecting ($u^{*}-r^{'}$, $g^{'}-z^{'}$) = ($-$1.0, 0.5), (0.5, 0.5), (3.0, 1.7), and (3.0, 5.0) \citep{2007ApJS..172..583H}.
However, contamination by extended sources becomes larger at $u^{*}-r^{'}>3$ which is most likely due to red galaxies at $z<0.4$ \citep[Figure 1 of ][]{2001AJ....122.1861S}. 
In contrast, we found that the $u^{*}-J$ versus $g^{'}-K_{\rm s}$ colour diagram (right panel of Figure \ref{fig:urgz-uJgKs-stellar-09502}) enables us to separate stars from galaxies with less contamination. 
The boundary of these regions can be drawn by the combination of straight lines connecting at 
($u^{*}-J$, $g^{'}-K_{\rm s}$)=($-$1.0,$-$1.4), (4.3, 2.1), (5.4, 3.5), and (9.0, 6.0).

\begin{figure}[t]
\begin{center}
  \includegraphics[width=35mm, angle=-90]{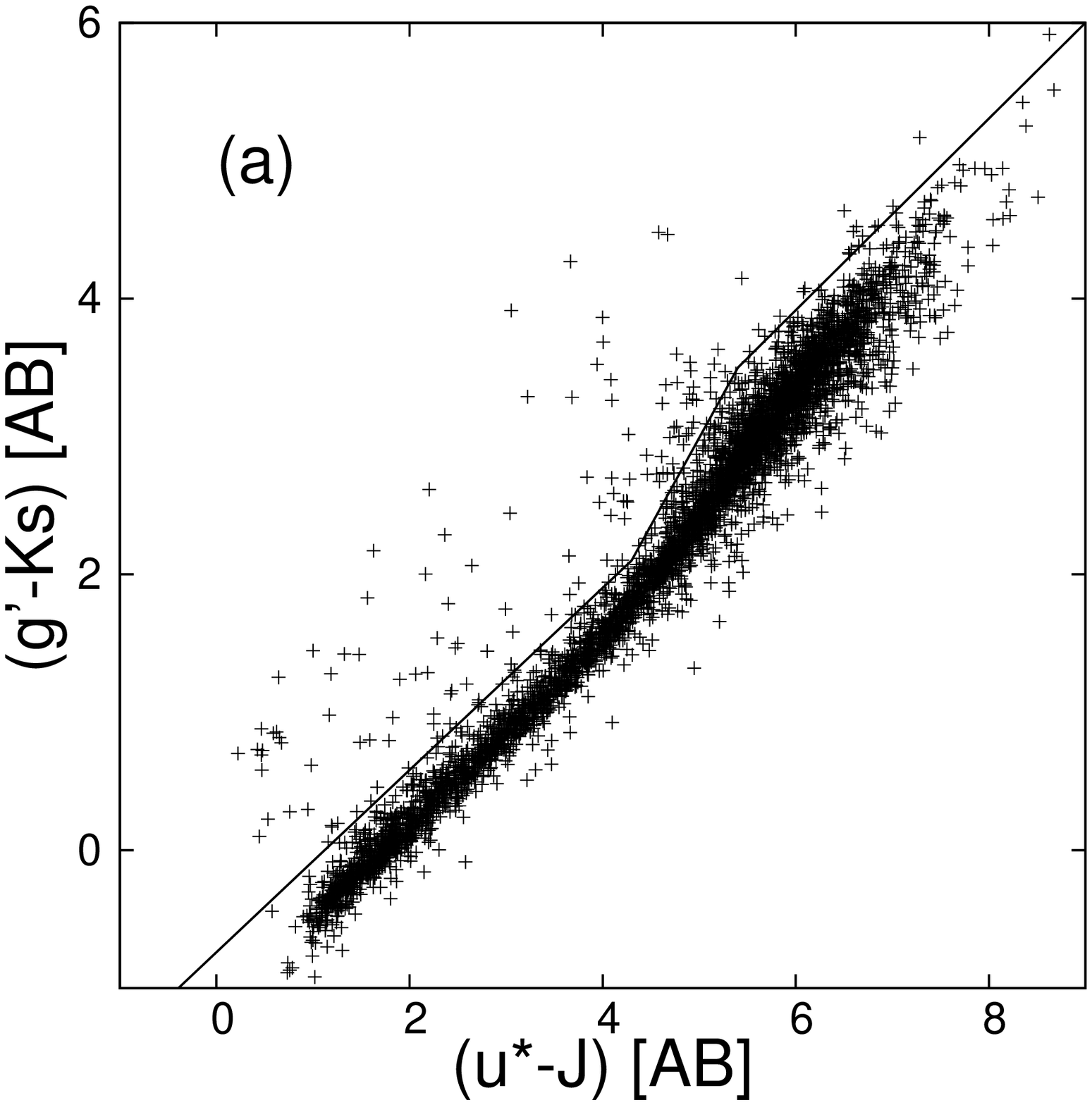}
\hspace{-1.2cm}
  \includegraphics[width=35mm, angle=-90]{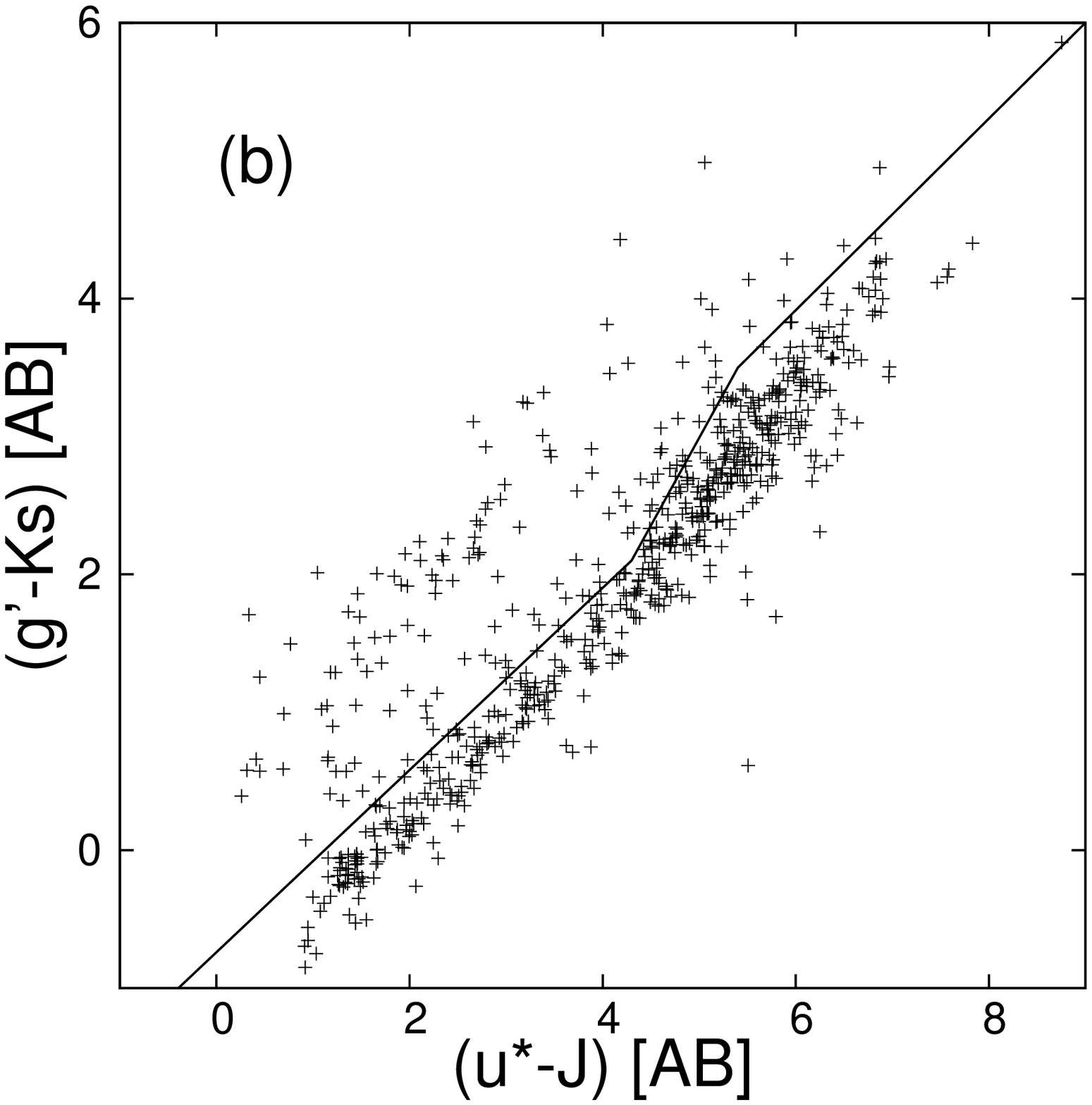}\\
  \includegraphics[width=35mm, angle=-90]{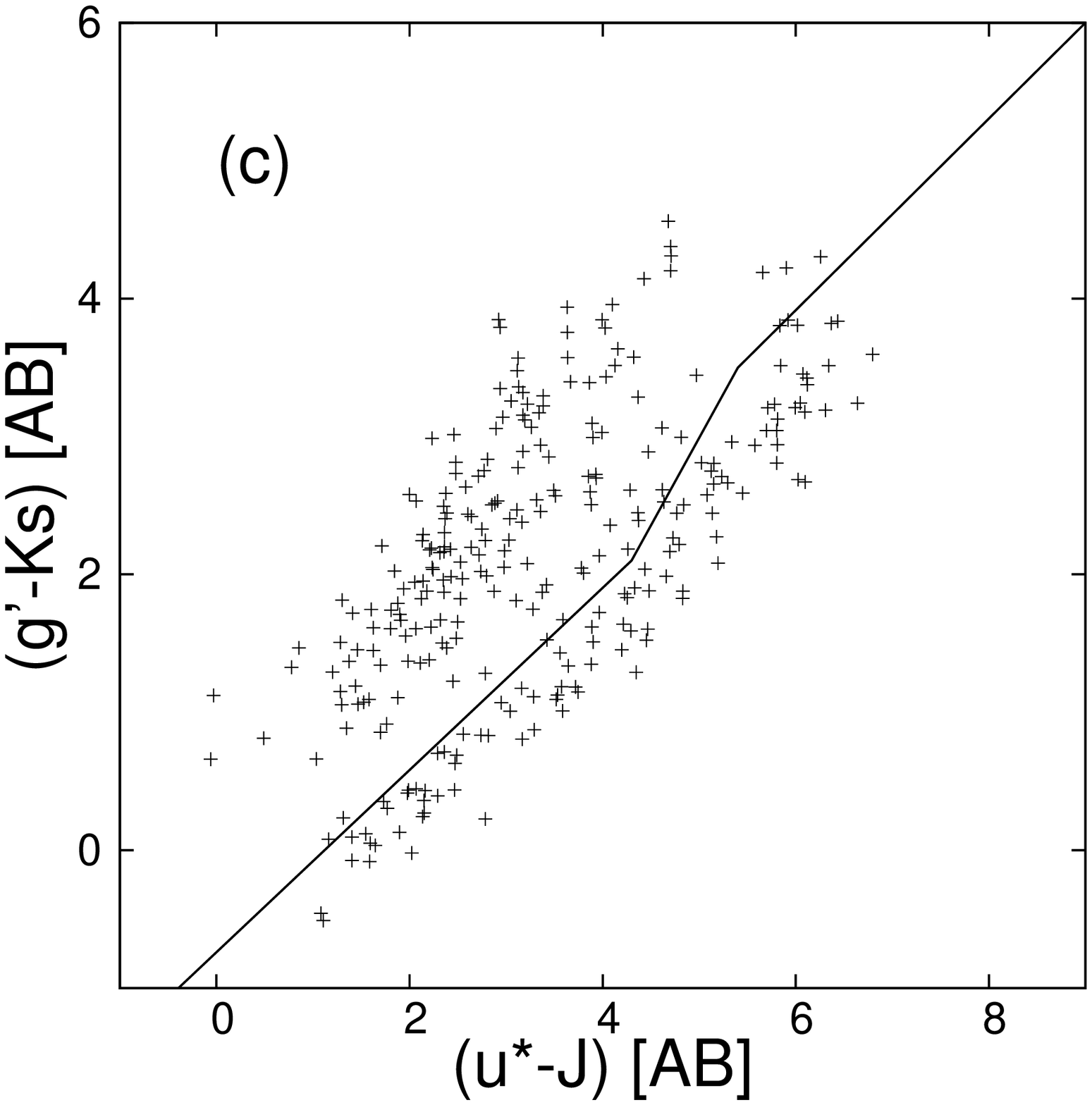}
\hspace{-1.2cm}
  \includegraphics[width=35mm, angle=-90]{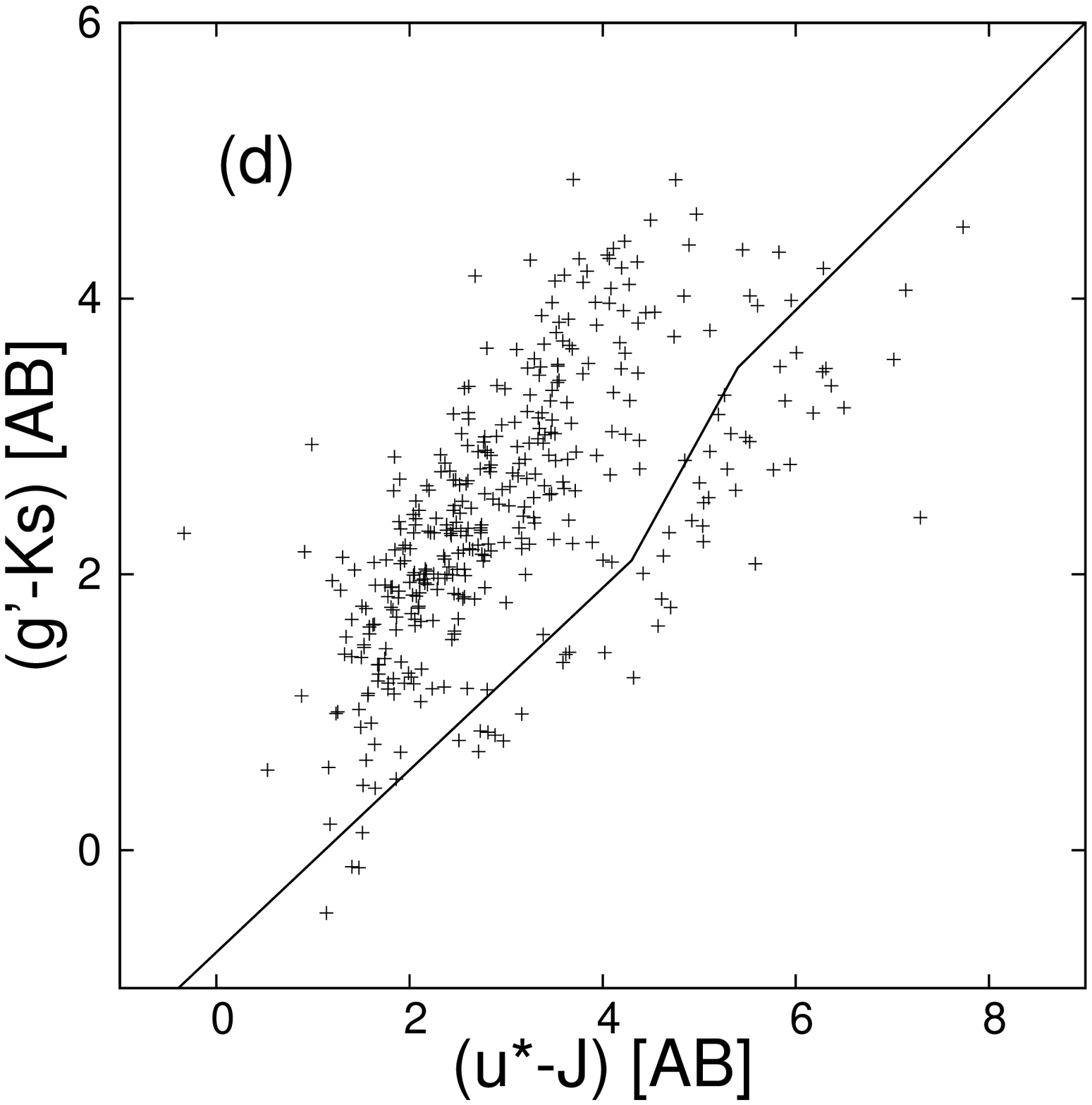}
  \caption{$u^{*}-J$ versus $g^{'}-K_{\rm s}$ colour-colour diagram for different stellarity index ranges at $z^{'}$-band. (a)--(d) panels represent 0.9 $<$ CLASS\_STAR $<$ 1.0, 0.8 $<$ CLASS\_STAR $<$ 0.9, 0.7 $<$ CLASS\_STAR $<$ 0.8, and 0.6 $<$ CLASS\_STAR $<$ 0.7, respectively.}
  \label{fig:uJgKs-stellar-1.0-0.9-0.8-0.7-0.6}
\end{center}
  \end{figure}

We also investigated colour dependence on the stellarity index ranges.
We plotted sources with 0.9 $<$ CLASS\_STAR $<$ 1.0, 0.8 $<$ CLASS\_STAR $<$ 0.9, 0.7 $<$ CLASS\_STAR $<$ 0.8, and 0.6 $<$ CLASS\_STAR $<$ 0.7 in the (a)--(d) panels of Figure \ref{fig:uJgKs-stellar-1.0-0.9-0.8-0.7-0.6}, respectively.   
The lines in all panels are the boundary shown in Figure \ref{fig:urgz-uJgKs-stellar-09502}.
The figure shows that most of the sources with CLASS\_STAR $>$ 0.8 in panels (a) and (b) fall under the boundary suggesting stars, while sources with CLASS\_STAR $<$ 0.7 in panel (d) are mainly located above the boundary indicating extended sources.
The sources at 0.7 $<$ CLASS\_STAR $<$ 0.8 seem to be a mixture of the two classes.
If we use only this colour criterion and stellarity index to separate stars from galaxies, quasars are likely to be misclassified as stars.
In order to avoid this, we excluded sources bluer than $u^{*}-g^{'}$ = 0.4 from the star candidates \citep{2000AJ....120.2615F}. 
In summary, the following criteria are for secure identification of stars; (1) $g^{'}-K_{\rm s}$ is bluer than the boundary in the $u^{*}-J$ versus $g^{'}-K_{\rm s}$ colour diagram, (2) CLASS\_STAR $>$ 0.8, and (3) $u^{*}-g^{'}>0.4$. 
5441 star candidates satisfy those criteria in our catalogue with a flag of "STAR". 
We note that the flagging of sources with at least one band non-detection in the $u^{*}$, $g^{'}$, $J$, and $K_{\rm s}$-bands is represented as "****" in the catalogue because the colour criteria cannot be used for them.

\section{PHOTOMETRIC REDSHIFT DERIVATION}
\label{sec:PHOTOMETRIC REDSHIFT}

\subsection{Photometric Redshift Computing with $Le Phare$} 
We computed photometric redshifts ($z_{\rm p}$) using $Le Phare$ code \citep{2006A&A...457..841I, 2007A&A...476..137A, 2009ApJ...690.1236I}.
We used 62 galaxy templates and the 154 star templates that we used for 
predicting $Y$-band magnitude as the photometric reference during the earlier data reduction (\S\ref{sec:wircam}).
The galaxy templates are empirical ones from \cite{1980ApJS...43..393C} with two starburst templates from \citep{1996ApJ...467...38K}. 
These templates are interpolated and adjusted to match the VIMOS VLT Deep Survey (VVDS) spectra better \citep{2007A&A...476..137A}. 
For galaxy templates of Scd and later spectral types, we considered the dust extinction reddening.
Since considerable changes in the extinction curve are expected from galaxy to galaxy, we used two extinction laws. 
We adopted an interstellar extinction law determined from starburst galaxies \citep{2000ApJ...533..682C} and the Small Magellanic Cloud extinction law \citep{1984A&A...132..389P} with varying $E(B-V)$ of 0.0, 0.05, 0.1, 0.15, 0.2, 0.3, 0.4, 0.5.
A broad absorption excess at 2175 $\AA$ (UV bump) sometimes necessary to explain rest-frame UV flux in some starburst galaxies \citep{2001A&A...380..425M}.
Therefore we allowed an additional UV bump at 2175 $\AA$ for Calzetti extinction law if it produces a smaller $\chi^2$. 	

Systematic offset are often found between the best fit SED templates and the observed apparent magnitudes in a given filter \citep{2006ApJ...651..791B, 2006A&A...457..841I, 2009ApJ...690.1236I}.
Uncertainties in the zero point calibration of the photometric data as well as imperfect knowledge of galaxy SEDs are responsible these offsets.
They may produce biases in the photometric redshift measurements.
In order to correct it, for each filter, $f$, we computed the systematic offset ($s^f$) between the photometry in a given filter and the photometry for the best SED fitting at the fixed spectroscopic redshift ($z_{\rm s}$) of the $AKARI$ detected galaxies with the option AUTO\_ADAPT in the $LePhare$ algorithm.
483 sources excluding Type 1 AGN taken with Keck/DEIMOS (Takagi et al., in prep.), MMT/Hectospec and WIYN/HYDRA \citep{2013ApJS..207...37S} are used as an input catalogue for the training and estimation of the $z_{\rm p}$ accuracy.
The all $z_{\rm s}$ of 483 galaxies were measured by using two or more emission lines or absorption lines.
Many objects with $z_{\rm s}$ distribute in the local universe at $z<1$ (453 sources $\sim$ 94\%) and only $\sim$ 6\% of the galaxy are at $z>1$.
Iteratively, we then searched for the best set of corrections that minimized the offsets.
Once found, the offset was applied to the SEDs when computing the photometric redshift for the entire catalogue.
The $z_{\rm p}$ is the redshift value which minimizes the merit function $\chi^2(z,T,A)$:
\begin{eqnarray}
\chi2=\Sigma^{N_f}_{f=1}\left(F^f_{obs}-A\times F^f_{pred}(z,T)10^{-0.4s^f} / \sigma^f_{obs}\right),
\end{eqnarray}
where $F^f_{pred}(z,T)$ is the predicted flux for a template $T$ at redshift $z$. $F^f_{obs}$ the observed flux, $\sigma^f_{obs}$ is the associated error, and $A$ is the normalization factor.
The grid spacing in redshift is $\delta z = 0.02$.
A redshift probability distribution function (PDFz) is computed for each object using the $\chi^2$ merit function, PDFz $\propto$ exp(--$\chi^2$($z$)/2) every grid space.
The best redshift is derived by parabolic interpolation of the PDF with galaxy library.
If a second peak is found in the PDFz with a height larger than 5\% of the main peak, the corresponding redshift is given as a second solution.
In addition to the best $\chi^2$ derived from the galaxy library (hearafter $\chi^2_{gal}$), a best $\chi^2$ computed using the stellar liberally is derived for each object (hearafter $\chi^2_{star}$).

\subsection{Photometric Redshift Accuracy}

\begin{figure}[t]
\begin{center}
  \includegraphics[width=60mm, angle=-90]{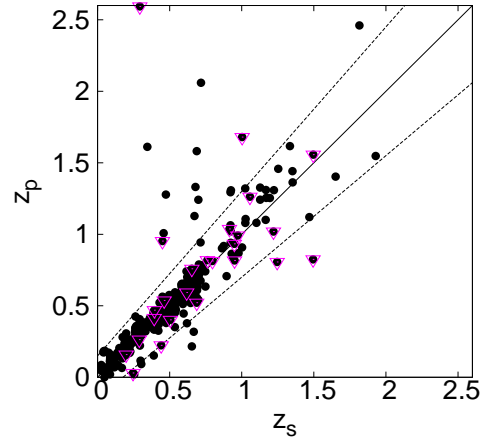}
  \caption{Comparison between the photometric redshift ($z_{\rm p}$) and the spectroscopic redshift ($z_{\rm s}$). Pink triangles represent objects for which we found a second solution in the PDFz. For the sources with single solution in the PDFz, an error of $z_{\rm p}$ is $\sigma_{\Delta z/(1+z)}$ = 0.032 and catastrophic error ($\Delta z$/$(1+z)>0.15$) rate is $\eta$ = 5.8\% at $z<1$, while $\sigma_{\Delta z/(1+z)}$ = 0.117 and  $\eta$ = 16.6\% for $z>1$. }
  \label{fig:zszp}
\end{center}
\end{figure}

We assess the photometric redshift accuracy by comparing photometric redshifts against spectroscopic redshifts.
Figure \ref{fig:zszp} shows the comparison between $z_{\rm p}$ with $z_{\rm s}$ from the 483 sources.
Objects, whose second solution in the PDFz are found, are shown with pink triangles.
Since $z_{\rm p}$ of those sources are not reliable, we excluded for the measurement of $z_{\rm p}$ accuracy.
We obtained an accuracy of $\sigma_{\Delta z/(1+z)}\sim$ 0.032 and a catastrophic error rate of $\eta=$ 5.8\% ($\frac{\Delta z}{1+z}>0.15$) at $z<1$, while $\sigma_{\Delta z/(1+z)}\sim$ of 0.117 and $\eta=$ 16.6\% at $z>1$.
The median offset between the photometric redshift and spectroscopic redshift is $\Delta z/(1+z)=$ 0.008 and 0.018 for $z<1$ and $z>1$, respectively.

\cite{2009A&A...500..981C} used the $LePhare$ with the CFHTLS Deep and Wide T0004 catalogues ($u^{*}, g^{'}, r^{'}, i^{'}, z^{'}$) to calculate $z_{\rm p}$ and measured the accuracy comparing with VVDS-deep and -wide, and DEEP2 (Davis et al 2003 2007) $z_{\rm s}$ samples.
They found a dispersion of 0.028-0.030 and an outlier rate of about 3-4\% in the range $17.5<i^{'}_{AB}<22.5$ at $0<z<2$.
In our catalogue, there are 382 sources with $z_{\rm s}$ in the range of $17.5<i^{'}_{AB}<22.5$, and ~97\% of them are $z_{\rm s}<1$. We compared the $z_{\rm p}$ with $z_{\rm s}$ of them and found an accuracy of $\sigma_{\Delta z/(1+z)}\sim$ 0.031 with a catastrophic error rate of $\eta=$ 3.5\%, which are comparable to the CFHTLS results.

Since evaluation of the $z_{\rm p}$ accuracy from the comparison with $z_{\rm s}$ is limited to specific range of magnitude and redshift, we use the 1$\sigma$ uncertainty in the derived $z_{\rm p}$ probability distribution to extend the uncertainty estimates over the full magnitude/redshift space.
For the estimate of the accuracy, we exclude stars.
In $\S$\ref{S-G separation}, we securely flagged star using ($u^{*}-J$) -- ($g^{'}-K_{\rm s}$) colour in addition to the stellarity index.
During SED fitting using $Le Phare$, we also calculated $\chi^2_{gal}$ and $\chi^2_{star}$ for each object.
If $\chi^2_{star}<\chi^2_{gal}$, it suggests that the object is more likely to be a star than a galaxy.
We compared the SED $\chi^2$ classification and colour with the stellarity index, and found that both classifications are excellently agreed with each other :  5381 out of 5441 star-flagged sources with the color-morphology classification method are classified as stars with the SED $\chi^{2}$ criterion.
It suggests that both criteria work well, and then we excluded stars flagged by either SED $\chi^2$ classification or the color-morphology classification for measure of the $z_{\rm p}$ accuracy below.

\begin{figure}[t]
\begin{center}
  \includegraphics[width=60mm, angle=-90]{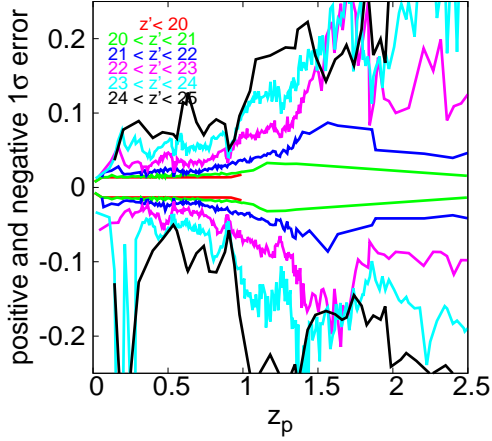}
  \caption{1$\sigma$ uncertainty for the $z_{\rm p}$ estimate as a function of photometric redshift in different $z^{'}$-band magnitude bin. Each value is computed with 100 galaxies per bin. The red, green, blue, pink, cyan, and black lines are $z^{'}<20$, $20<z^{'}<21$, $21<z^{'}<22$, $22<z^{'}<23$, $23<z^{'}<24$, and $24<z^{'}<25$, respectively.}
  \label{fig:1sigma}
\end{center}
\end{figure}

Following \cite{2006A&A...457..841I}, we plotted the 1$\sigma$ negative and positive uncertainties derived from the PDFz as a function of redshift and $z^{'}$-band magnitude (Figure \ref{fig:1sigma}).
The uncertainties are derived from at 68\% of the probability distribution.
This figure shows that the accuracy is inevitably degraded for fainter galaxies at all redshift and the $z_{\rm p}$ have significantly higher uncertainty at $z_{\rm p}$ \hspace{0.3em}\raisebox{0.3ex}{$>$}\hspace{-0.6em}\raisebox{-.7ex}{$\sim$}\hspace{0.3em} 1.
From $z_{\rm p}>0$ up to $z_{\rm p}$ = 1, the 1$\sigma$ errors  do not dependent significantly on the redshift at $z^{'}<$ 23. 
With $z^{'}$ = 23 -- 24, the accuracy is $\sim$ 0.7, which is consistent with the accuracy estimated from the $z_{\rm p}$ -- $z_{\rm s}$ comparison (see Figure \ref{fig:zszp}).
At $z_{\rm p}>1$, the accuracy with $z^{'}<$ 22 is continuously small, while the accuracy for fainter sources goes high.
Therefore, our photometric redshifts are highly accurate up to at least $z_{\rm p}<1$ with $z^{'}<24$ or $z_{\rm p}<2.5$ with $z^{'}<$ 22.

\begin{figure}[t]
\begin{center}
  \includegraphics[width=60mm, angle=-90]{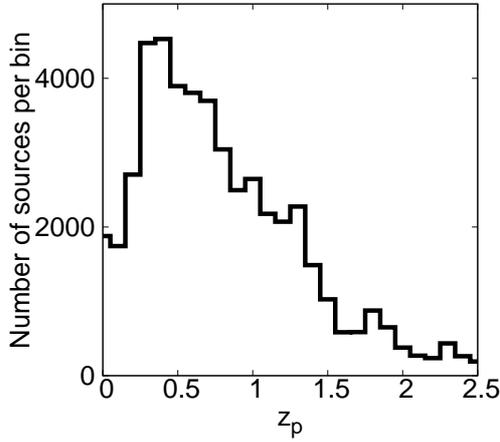}
  \caption{Photometric redshift distribution between $0<z_{\rm p}<2.5$.}
  \label{fig:hist-zp}
\end{center}
\end{figure}

Figure \ref{fig:hist-zp}  shows the photometric redshift distribution for all objects classified as galaxy with at least five band detections. 
The peak of the distribution is located at redshifts between at $z\sim$ 0.3 -- 0.4, and the number of sources monotonically decreases with increasing redshift. 

\section{$AKARI$ SOURCES}
\cite{2013A&A...559A.132M} constructed a revised near to mid infrared catalogue for the $AKARI$ NEP-Deep survey. \if by combining revised $AKARI$ near-infrared (NIR) a sub-catalogue ($N2$--$N4$) and mid-infrared (MIR) sub-catalogue ($S7$--$L24$).\fi
They compared the $AKARI$ NEP-Deep catalogue with our final catalogue to identify optical counterparts.
In total, 23$\hspace{0.1em}$345 $AKARI$ sources have their optical counterparts, and 8916 of them have mid-infrared (any of $S7$ -- $L24$) fluxes.
In this section, we will focus on the sources detected by $AKARI$ with optical counterparts and demonstrate the properties of the $AKARI$ sources.

\subsection{$g^{'}z^{'}K_{\rm s}$ Diagram}
\label{sec:gzk}
\cite{2004ApJ...617..746D} developed a highly popular technique that allows both the selection and classification of $1.4<z<2.5$ star-forming galaxies, passively evolved galaxies and stars using a simple $BzK_{\rm s}$ colour-colour diagram.
In the $BzK_{\rm s}$ technique, high redshift objects are uniquely located in regions of the $BzK_{\rm s}$ diagram.
The validity of this technique has been supported by many authors. 
Original paper of \cite{2004ApJ...617..746D} used the VLT Bessel $B$ band filter, the VLT Gunn $z$ filter, and the ISAAC $K_{\rm s}$ band filter.
Although the $z^{'}$ and $K_{\rm s}$-band filters of CFHT are similar to those Daddi et al. used, the CFHT MegaCam $g^{'}$-band filter has offset from the $B$ band filter.
\cite{2013arXiv1307.6094L} considered the positions of the \cite{2004ApJ...617..746D} models in the $g^{'}z^{'}K_{\rm s}$ colour-colour diagram to modify the $BzK_{\rm s}$ technique for use with MegaCam and WIRCam $g^{'}z^{'}K_{\rm s}$ filters. 
They found that even the $g^{'}z^{'}K_{\rm s}$ and $BzK_{\rm s}$ filter systems are different, the main qualitative features of the colour-colour diagram of models are similar enough.
Hence, $g^{'}z^{'}K_{\rm s}$ diagram can be used for classification of $1.4<z<2.5$ star-forming galaxies, passively evolved galaxies and stars.
\cite{2013arXiv1307.6094L} defined new colour-cats to distinguish those properties as 
\begin{eqnarray}
(z^{'}-K_{\rm s}) - 1.27(g^{'}-z^{'})&=& -0.022\\
(z^{'}-K_{\rm s}) - 1.27(g^{'}-z^{'})&=& -0.022\cap(z^{'}-K_{\rm s}) =2.55\\
(z^{'}-K_{\rm s})-0.45(g^{'}-z^{'})&=& -0.57.
\end{eqnarray}
As the same as $BzK_{\rm s}$ technique, high redshift star-forming galaxies are located in the left of the Eq. (2) and stars are separated by Eq. (4) in the $g^{'}z^{'}K_{\rm s}$ diagram. 

\begin{figure}[t]
\begin{center}
   \includegraphics[width=35mm, angle=-90]{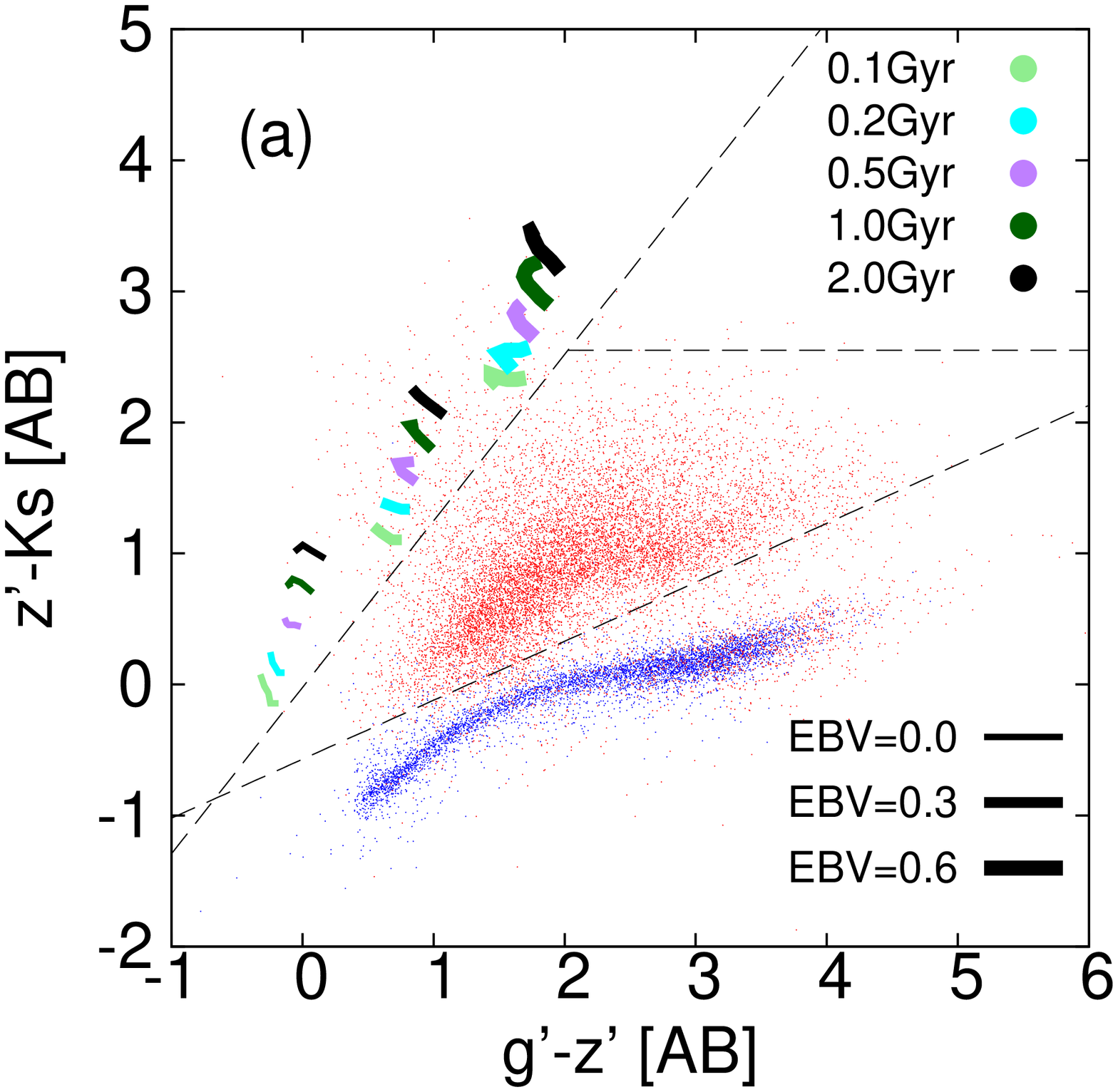}
\hspace{-1.2cm}
    \includegraphics[width=35mm, angle=-90]{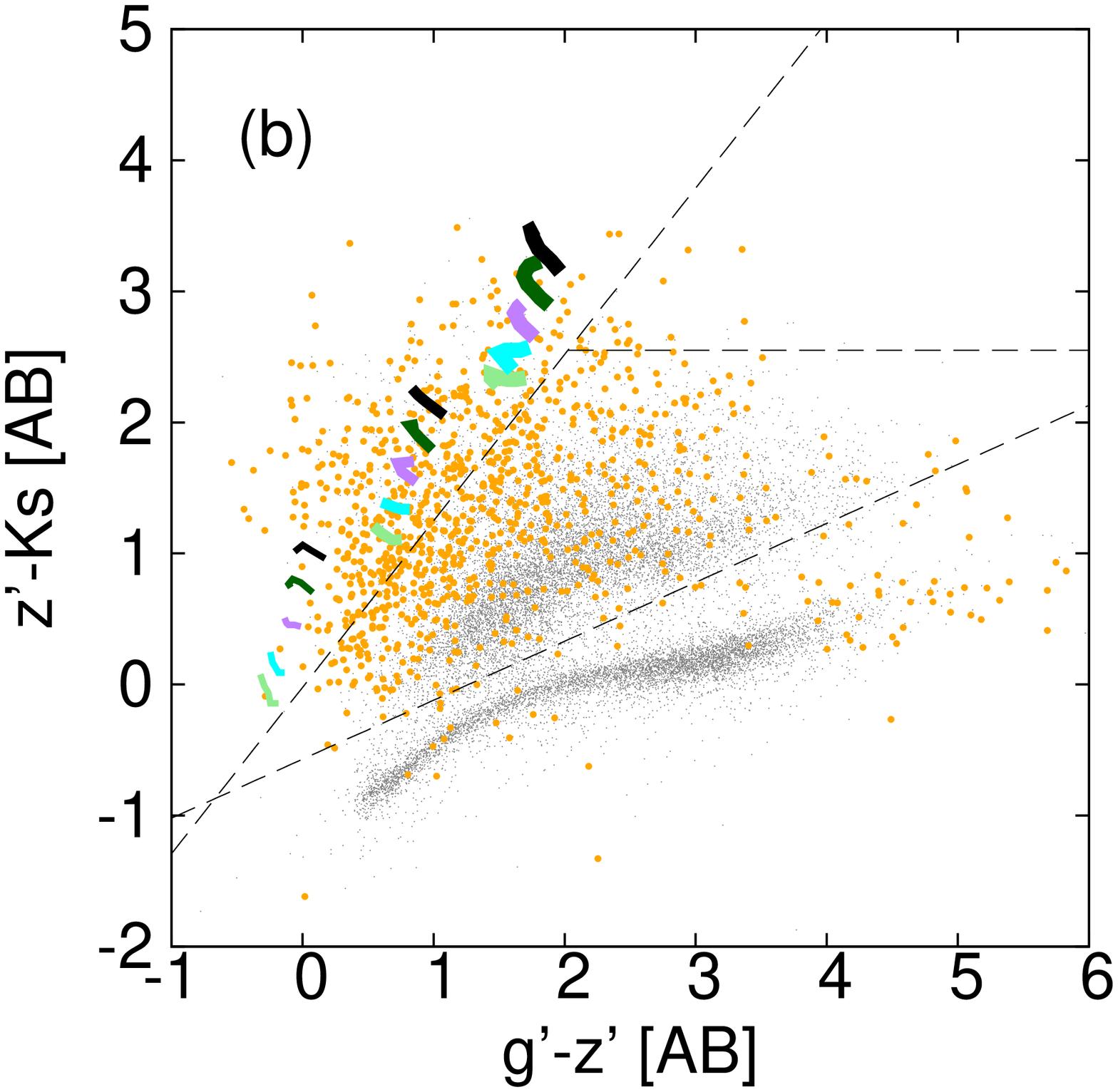}
  \caption{$g^{'}z^{'}K_{\rm s}$ diagram for $AKARI$ sources with optical counterparts. (Left) blue dots represent stars identified in $\S$\ref{S-G separation}, and red dots indicate galaxies with $z_{\rm p}<1.4$.
The dashed lines are the borders given by Eq. (2) -- (4).
\cite{2003MNRAS.344.1000B} stellar population synthesis models for $g^{'}z^{'}K_{\rm s}$ colour--colour diagram with $1.4<z_{\rm p}<2.5$ which \cite{2013arXiv1307.6094L} showed are over-plotted as comparison.
Ages in Gyr are shown with different colours while different $E(B - V)$ are represented with different thickness of lines.
(Right) $1.4<z_{\rm p}<2.5$ $AKARI$ sources are plotted in yellow. The $z_{\rm p}<1.4$ sources shown in the left panel are also plotted but in gray.
}
  \label{fig:gzkslgt4}
\end{center}
\end{figure}

We first plotted the $z_{\rm p} <1.4$ $AKARI$ sources (left panel of Figure \ref{fig:gzkslgt4}).
Almost all stars ($\sim$ 98.9\%) classified in $\S$\ref{S-G separation} are lying under the Eq. (4), while extended galaxies are distributed enclosed area by the three equations.
It suggests that the criteria of star--galaxy classification discussed in $\S$\ref{S-G separation} work well.
On the other hand, we plotted galaxies at $1.4<z_{\rm p}<2.5$ in the $g^{'}z^{'}K_{\rm s}$ diagram in right panel of Figure \ref{fig:gzkslgt4}.
All points in the left panel are plotted in gray.
We found that the $1.4<z_{\rm p}<2.5$ galaxies are located at upper-left of galaxies of $z_{\rm p} <1.4$ systematically although more than 50\% of $1.4<z_{\rm p}<2.5$ galaxies are scattered into $z_{\rm p} <1.4$ region.
This is probably a consequence of the increased scatter of the photometric redshift higher than $z_{\rm p}>1$.
When we use the $g^{'}z^{'}K_{\rm s}$ diagram, we can distinguish 759 $AKARI$ sources ($\sim$ 3.2\%) as high-$z$ galaxies independent of $z_{\rm p}$ with large uncertainty at $z_{\rm p}>1$.
The average $z^{'}$ magnitude of the high-$z$ galaxies of $1.4<z_{\rm p}$ is 23.2 ($\pm$ 0.4) mag, which is consistent with the $z^{'}$-band 50\% completeness, suggesting that it is possible there are $AKARI$ sources of high-$z$ star-forming galaxies without optical counterparts.  
We also found that only less than 0.1\% of $AKARI$ sources come into the passive galaxies' area where is above the Eq. (3).

Seen from the $g^{'}z^{'}K_{\rm s}$ diagram, the $AKARI$ sample is dominated by sources at $z<1.4$, and $<$ 5\% of the $AKARI$ sources with optical counterparts are located at the high-$z$ universe and most of them seem to be star-forming active galaxies.

\subsection{Mid Infrared Sources}
\label{sec:AKARI}

\begin{figure}[htbp]
\begin{center}
  \includegraphics[width=60mm, angle=-90]{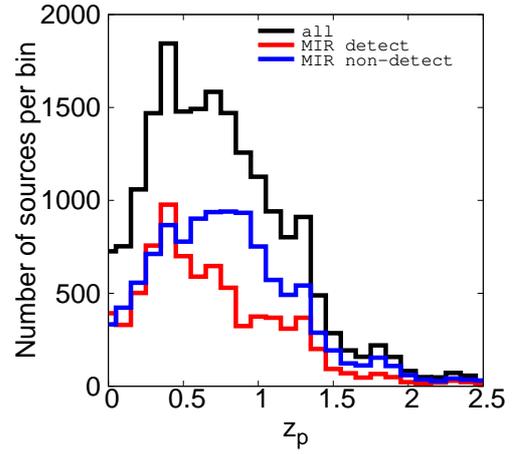}
  \caption{Histograms of $z_{\rm p}$ for $AKARI$ sources at $0<z_{\rm p}<2.5$. Sources flagged as stars in $\S$\ref{S-G separation} are excluded. Black, red and blue lines represent all sources, sources with and without MIR detections, respectively. }
  \label{fig:histwwoAKARI}
\end{center}
\end{figure}

We compared $z_{\rm p}$ between sources with detections and non-detections in $AKARI$ MIR bands.
We divided $AKARI$ sources into two groups, one is detected in any of MIR bands ($S7$--$L24$) and the other is not. 
After excluding star-flagged sources, MIR detected and non-detected groups include 8085 sources and 11$\hspace{0.1em}$616 sources, respectively.
Figure \ref{fig:histwwoAKARI} shows $z_{\rm p}$ histograms for each group.
This figure shows that MIR non-detected group is distributed at higher redshift systematically compared with MIR detected group.
The distribution of MIR detected group has sharp peak around $z\sim 0.5$ and suddenly decreases at higher redshift, while MIR non-detected sources have gentle peak around $z_{\rm p} \sim 0.5 - 1$.
This tendencies are explained by the difference of sensitivities of $AKARI$ near and mid infrared detectors.
Since mid-infrared band sensitivities are relatively poorer than those of near-infrared bands, there is a bias that MIR bands detect lower-redshift bright sources.

\begin{figure}[htbp]
\begin{center}
  \includegraphics[width=35mm, angle=-90]{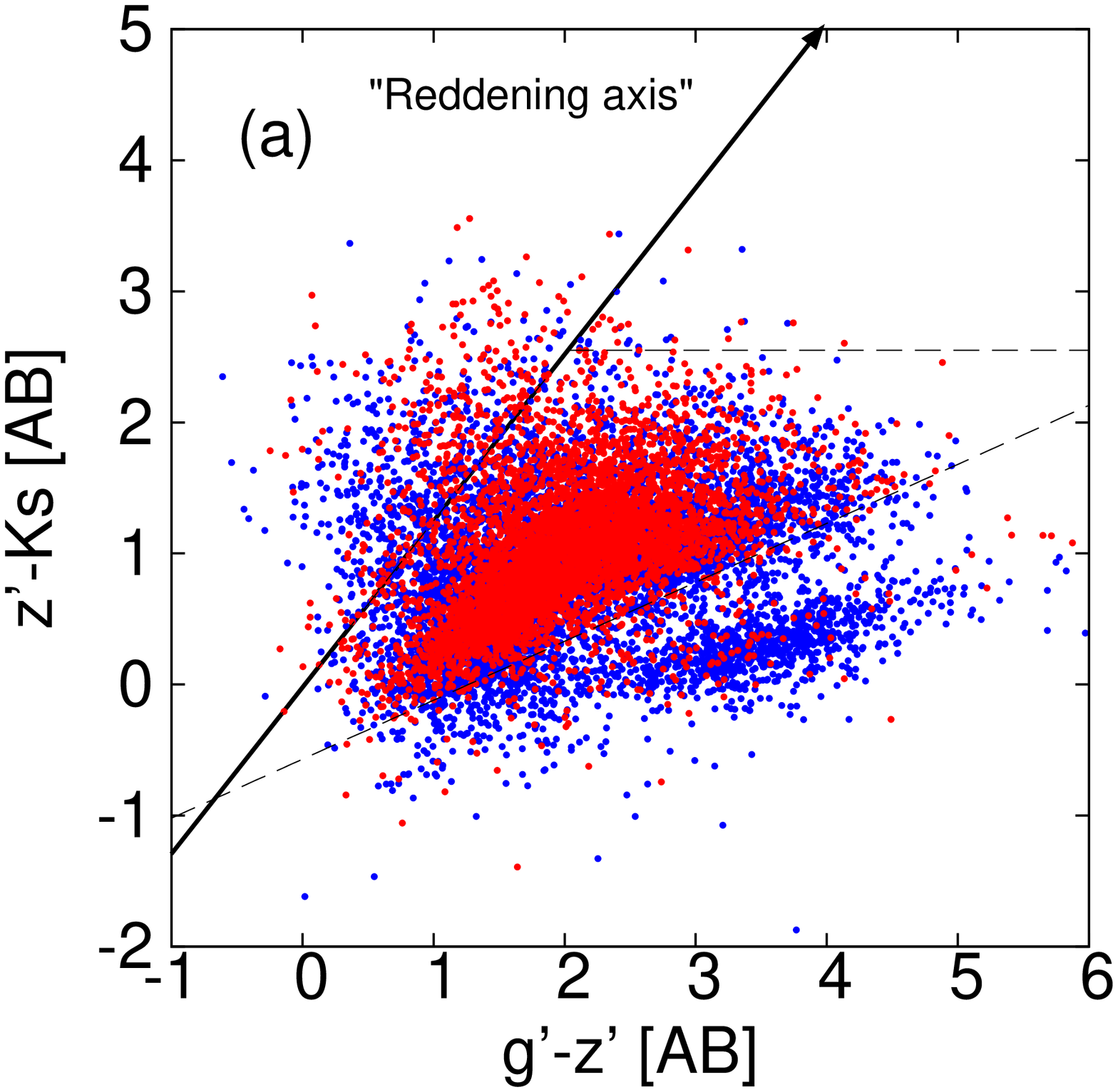}
\hspace{-1.2cm}
  \includegraphics[width=35mm, angle=-90]{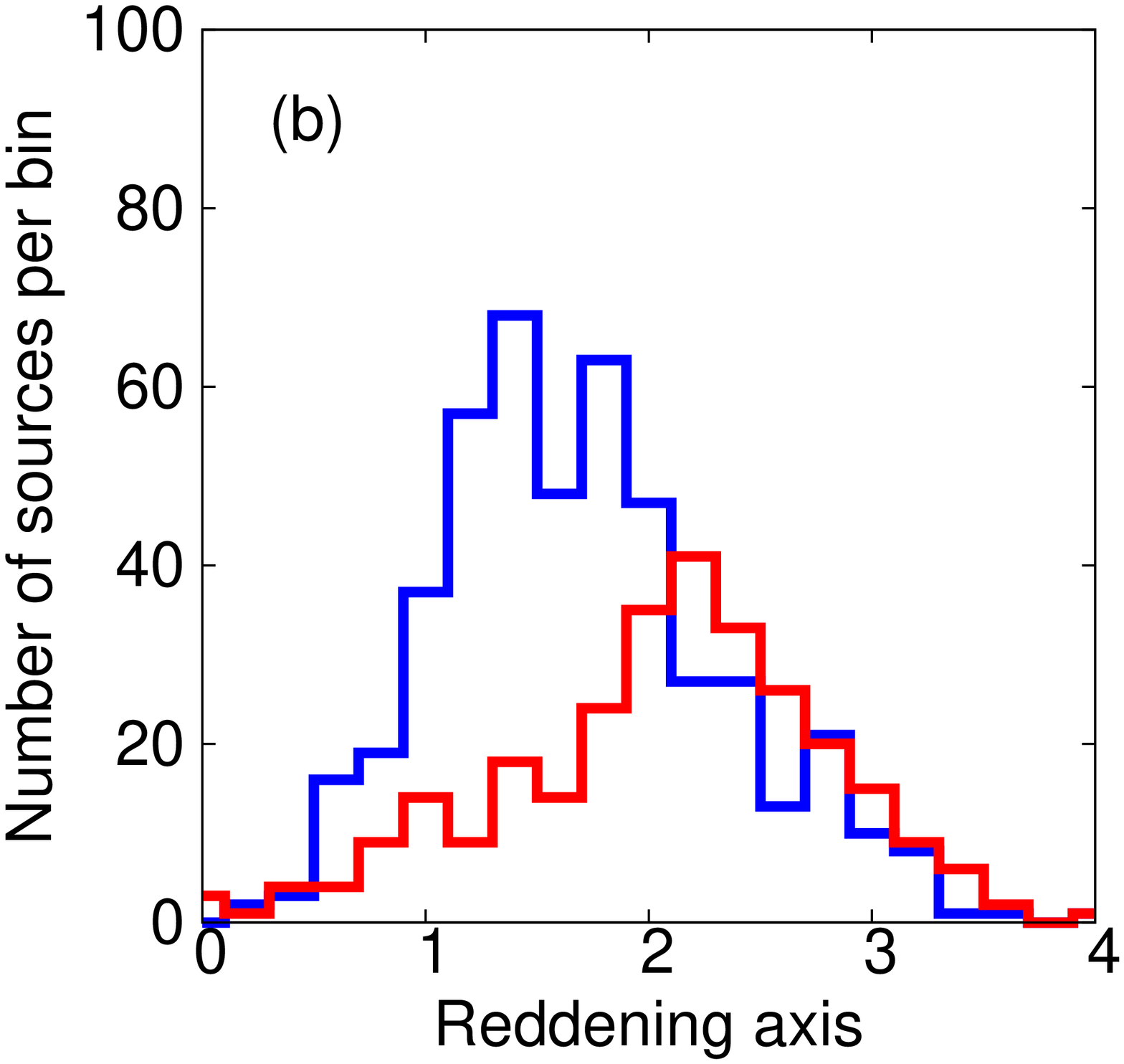}
  \caption{Left: $g^{'}z^{'}K_{\rm s}$ diagram for sources with and without detections in $AKARI$ MIR bands.
 The colour of indices are the same as Figure \ref{fig:histwwoAKARI}. 
 The solid thick arrow is the axis defined along the line of Eq. (2) named "Reddening axis" (see text).
   Right: distribution along the "Reddening axis" for $AKARI$ high-$z$ star-forming sources above the Eq. (2) in the left panel.
   The values along the "Reddening axis" are calculated by coordinate transformation from $g^{'}z^{'}K_{\rm s}$ diagram.}
  \label{fig:reddening}
\end{center}
\end{figure}

Finally we plotted the $AKARI$ sources in the $g^{'}z^{'}K_{\rm s}$ plane in Figure \ref{fig:reddening}.
This figure shows that some sources with non-detections in $AKARI$ MIR bands ($\sim$ 22\%) are plotted in the stellar region under the Eq. (4), while stellar contamination to the sources with detections in the MIR bands are negligible small ($\sim$ 4.0\%).
Very similar fraction of sources with and without MIR detections ($\sim$ 5.3\% and $\sim$ 6.4\%, respectively) are distributed in the $1.4<z<2.5$ star-forming galaxy region.

For these sources located in the $1.4<z<2.5$ star-forming galaxy region, MIR detected sources seem to be systematically redder along the line of Eq. (2) compared with MIR non-detected group.
Since galaxies become redder along the line of Eq. (2) depending on the $E(B-V)$ values in \cite{2003MNRAS.344.1000B} model as shown in Figure \ref{fig:gzkslgt4}, 
here we defined an axis along the line of Eq. (2) as "Reddening axis". 
The difference between MIR detected and non-detected groups can be more clearly seen along the "Reddening axis" as shown in right panel of Figure \ref{fig:reddening}.
Although the colour can be redder both from the old age and the star formation along the "Reddening axis", the effect from the age is significantly small.
Therefore, the redder colour of MIR detected sources probably indicates that they are dust attenuated star-forming galaxies.

In conclusion, when we select sources detected in $AKARI$ MIR bands, then dusty, star-forming galaxies at around $z$\hspace{0.3em}\raisebox{0.3ex}{$<$}\hspace{-0.6em}\raisebox{-.7ex}{$\sim$}\hspace{0.3em}0.5 are preferentially-chosen with less stellar contamination.

\section{SUMMARY and CONCLUSION}
We obtained $g^{'}$, $r^{'}$, $i^{'}$, $z^{'}$, $Y$, $J$, and $K_{\rm s}$-band images with MegaCam and WIRCam at CFHT, and cross-matched each of all the band catalogues with existing $u^{*}$-band catalogue to construct an optical -- near infrared 8-band merged photometric catalogue covering the $AKARI$ NEP-Deep field.
The 4$\sigma$ detection limits over an 1 arcsec radius aperture are 26.7, 25.9, 25.1, 24.1, 23.4, 23.0, and 22.7 AB mag for $g^{'}$, $r^{'}$, $i^{'}$, $z^{'}$, $Y$, $J$, and $K_{\rm s}$-bands, respectively.
The astrometry of the band-merged catalogue is offset by 0.013 arcsec to the 2MASS with an RMS offset of 0.32 arcsec, while the RMS offsets between $z^{'}$-band and the other bands are 0.08--0.11 arcsec, but that between $z^{'}$ and $u^{*}$-bands is worse (0.18 arcsec) because of a slightly degraded PSF in $u^{*}$-band.
We securely distinguished stars from galaxies with ($u^{*}-J$) versus ($g^{'}-K_{\rm s}$) colour-colour diagram with CLASS\_STAR of $<$ 0.8. 
We derived photometric redshifts based of a SED template fitting procedure using $LePhare$. 
The errors comparison with $z_{\rm s}$ for our spectroscopic sub-sample reveals the dispersion of $\sigma_{\Delta z/(1+z)}\sim0.032$ with $\eta=5.8$\% catastrophic failures at $z_{\rm p}<1$.
We extrapolate this result to fainter magnitudes using the 1$\sigma$ uncertainties in the $z_{\rm p}$ probability distribution functions.
At $z_{\rm p}<1$, we estimate a $z_{\rm p}$ accuracy of $\sigma_{\Delta z} = 0.7$ with $z^{'}<24$, while the accuracy is strongly degraded at $z_{\rm p}>1$ or $z^{'}>24$.
We investigated properties of $AKARI$ sources with optical counterparts from our catalogue.
We plotted the sources with $g^{'}z^{'}K_{\rm s}$ diagram and found that most of the $AKARI$ sources are located in the low-$z$ ($z<1.4$) and $<$ 5\% of them are classified as high-$z$ ($1.4<z<2.5$) star-forming galaxies. 
Among the high-$z$ star-forming galaxies, MIR detected sources seems to be affected stronger dust extinction compared with sources with non-detections in $AKARI$ MIR bands.
Our final catalogue contains 85$\hspace{0.1em}$797 sources with $z_{\rm p}$, which are detected in $z^{'}$-band and at least one of the other bands. 
This optical to near-infrared photometric data with photometric redshift in the $AKARI$ NEP-Deep field derived here are crucial to study star-formation history at the era of active universe ($z\sim1$) using multi-wavelength data sets ($Chandra$, $GALEX$, $AKARI$, $Herschel$, $SCUBA$-2, and $WSRT$).

\begin{acknowledgements} 

This work is supported by the Japan Society for the Promotion of Science (JSPS; grant number 23244040) and grant NSC 100-2112-M-001-001-MY3 (Y.O.).
MI acknowledges the support from a grant No. 2008-0060644 of CRI/NRFK/MSIP of Korea.
This work is based on observations obtained with MegaPrime/MegaCam, a joint project of CFHT and CEA/DAPNIA, and WIRCam, a joint project of CFHT, Taiwan, Korea, Canada, France, at the Canada-France-Hawaii Telescope (CFHT) which is operated by the National Research Council (NRC) of Canada, the Institute National des Sciences de l'Univers of the Centre National de la Recherche Scientifique of France, and the University of Hawaii.
S. Foucaud (National Taiwan Normal University) is acknowledged for this modified SWarp in Section \ref{sec:OBSERVATION and DATA REDUCTION}

\end{acknowledgements}

\clearpage
\twocolumn[{
\appendix
\section{CATALOGUE FORMAT}
We give a short description of the columns in the catalogue in the following.\\
\noindent
Column (1) : Identification number. \\
\noindent
Columns (2) and (3) : the {\it J}2000 right ascension and the declination, in degrees. 
The coordinates are based on the $z^{'}$-band astrometry.\\
\noindent
Columns (4) and (5) : the AB magnitudes of the sources in the $z^{'}$-band, and $z^{'}$-band magnitude error, respectively.\\
\noindent
Columns (6), (9), (12), (15), (18), (21), and (24) : the AB magnitudes of the other bands ($u^{*}$, $g^{'}$, $r^{'}$, $i^{'}$, $Y$, $J$, and $K_{\rm s}$-bands).\\
\noindent
Columns (7), (10), (13), (16), (19), (22), and (25) : the magnitude errors for the other bands. \\
\noindent
Columns (8), (11), (14), (17), (20), (23), and (26) : the cross-matching flags. $^{"}$0$^{"}$: a unique counterpart is found for the $z^{'}$-band sources. $^{"}$1$^{"}$: multiple counterparts are found for the $z^{'}$-band sources. The magnitude is the sum of all the candidate fluxes. $^{"}$2$^{"}$: multiple $z^{'}$-band sources have one common counterpart in other band. 
If counterparts are not found, $^{"}-$99.00$^{"}$ is inserted as the dummy values into magnitude and error columns of the no-counterpart bands with $^{"}-$1$^{"}$ flag.\\
\noindent
Column (27) : the CLASS\_STAR at $z^{'}$-band image.\\
\noindent
Column (28) : flags of stars based on colour-colour diagram. $^{"}$STAR$^{"}$ is identified as a star and $^{"}$****$^{"}$, is the others.\\
\noindent
Columns (29) and (30): spectroscopic redshift and flag measured by Takagi et al. (in prep.) using Keck/DEIMOS.
$^{"}$5$^{"}$ : secure redshift with multiple lines, including spectroscopically resolved [OII] doublet.
$^{"}$3$^{"}$ : redshift from absorption lines only.
$^{"}$2$^{"}$ : redshift from single line, probably [OII].
$^{"}$1$^{"}$ : single line with less confidence.
$^{"}-$1$^{"}$ : problem in data.
$^{"}-$3$^{"}$ : redshift from absorption lines only, with problems in data.
$^{"}-$5$^{"}$ : secure redshift, but problems in data.
$^{"}-$99$^{"}$ : no Keck/DEIMOS spectral data.\\
\noindent
Columns (31),(32) and (33): spectroscopic redshift, flag, and classification by \cite{2013ApJS..207...37S}, using MMT/Hectospec and WIYN/HYDRA. 
$^{"}$4.0$^{"}$ : the redshift is clearly identified using more than two significant spectral features.
$^{"}$3.0$^{"}$ : the redshift is very likely identified. Multiple spectral features are used, but some of the features have low signal-to-noise ratio. 
$^{"}$2.5$^{"}$ : objects which seem to be between flags 2 and 3.
$^{"}$2.0$^{"}$ : the redshift is identified based on single, weak spectral features (including continuum break). Probable, but unreliable. 
$^{"}$1.5$^{"}$ : the spectral features are very weak, might be real, but are not reliable.
$^{"}-$99$^{"}$ : no Keck/DEIMOS spectral data.\\
There are 5 classifications; GALAXY, STAR, TYPE1, TYPE2, and UNKNOWN.\\
\noindent
Column (34) : photometric redshift for galaxy measured by SED fitting ($\S$\ref{sec:PHOTOMETRIC REDSHIFT}). Star-flagged sources in columns (28) and (33) have a $z_{\rm p}$ value of 0.000.\\
}
Column(35) : Classes (star/galaxy) from SED fitting $\chi^2$ values.]

\setcounter{table}{6} 

\begin{landscape}
\begin{table}[htpb]
\caption{Optical -- near-infrared catalogue at $AKARI$ NEP field. The full, electric catalogue of this table is available at the CDS, contains the following information.
}
\scalebox{0.97}{ 
 \begin{tabular}{rrrrrrrcrrcrrcrrcr}
\hline
\hline
\multicolumn{1}{c}{ID} 
& \multicolumn{1}{c}{RA} 
& \multicolumn{1}{c}{Dec} 
&\multicolumn{3}{c}{$-u^{*}-$}
&\multicolumn{3}{c}{$-g^{'}-$}
&\multicolumn{3}{c}{$-r^{'}-$}
&\multicolumn{3}{c}{$-i^{'}-$}
&\multicolumn{2}{c}{$-z^{'}-$}
&\\
\multicolumn{1}{c}{} 
&\multicolumn{1}{c}{[deg]} 
&\multicolumn{1}{c}{[deg]} 
&\multicolumn{1}{c}{mag}
&\multicolumn{1}{c}{err}
&\multicolumn{1}{c}{flag}
&\multicolumn{1}{c}{mag}
&\multicolumn{1}{c}{err}
&\multicolumn{1}{c}{flag}
&\multicolumn{1}{c}{mag}
&\multicolumn{1}{c}{err}
&\multicolumn{1}{c}{flag}
&\multicolumn{1}{c}{mag}
&\multicolumn{1}{c}{err}
&\multicolumn{1}{c}{flag}
&\multicolumn{1}{c}{mag}
&\multicolumn{1}{c}{err}
&\\
\multicolumn{1}{c}{(1)} 
&\multicolumn{1}{c}{(2)}
&\multicolumn{1}{c}{(3)}
&\multicolumn{1}{c}{(4)}
&\multicolumn{1}{c}{(5)}
&\multicolumn{1}{c}{(6)}
&\multicolumn{1}{c}{(7)}
&\multicolumn{1}{c}{(8)}
&\multicolumn{1}{c}{(9)}
&\multicolumn{1}{c}{(10)}
&\multicolumn{1}{c}{(11)}
&\multicolumn{1}{c}{(12)}
&\multicolumn{1}{c}{(13)}
&\multicolumn{1}{c}{(14)}
&\multicolumn{1}{c}{(15)}
&\multicolumn{1}{c}{(16)}
&\multicolumn{1}{c}{(17)}
&\\
\hline
3 & 269.4974 & 66.0392 &  -99.000 & -99.000 & -1 & -99.000 & -99.000 & -1 & 23.494 & 0.109 & 0 & -99.000 & -99.000 & -1 & 23.388 & 0.172 &\\
721 & 268.4390 & 66.0463 &  -99.000 & -99.000 & -1 & 20.451& 0.003 & 0 & 19.591 & 0.003 & 0 & 19.118 & 0.003 & 0 & 18.669 & 0.005 &\\
5414 & 268.7602 & 66.1114 &  24.319 & 0.100 & 1 & 24.017 & 0.023 & 0 & 23.908 & 0.034 & 0 & 23.653 & 0.044 & 0 & 22.603 & 0.064 &\\
15159 & 268.8986 & 66.2231 &  26.641 & 0.198 & 0 & 23.803 & 0.016 & 0 & 22.618 & 0.011 & 0 & 21.671 & 0.008 & 0 & 21.041 & 0.013 &\\
21777 & 268.8178 & 66.3065 &  -99.000 & -99.000 & -1 & 24.527 & 0.040 & 0 & -99.000 & -99.000 & -1 & 23.466 & 0.071 & 0 & 23.120 & 0.102 &\\
44667 & 268.5258 & 66.5782 &  24.221 & 0.038 & 0 & 23.633 & 0.024 & 0 & 23.303 & 0.037 & 0 & 23.075 & 0.049 & 2 & 22.480 & 0.080 &\\
69588 & 268.9472 & 66.8808 &  -99.000 & -99.000 & -1 & 24.132 & 0.023 & 0 & 22.745 & 0.013 & 0 & 21.700 & 0.010 & 0 & 21.191 & 0.018 &\\
79677 & 268.7019 & 66.9967 &  21.614 & 0.007 & 0 & 19.801 & 0.001 & 0 & 18.765 & 0.001 & 0 & 18.225 & 0.001 & 0 & 17.748 & 0.002 &\\
84181 & 270.2342 & 67.0469 &  -99.000 & -99.000 & -1 & 22.954 & 0.011 & 0 & 22.351 & 0.012 & 0 & 21.995 & 0.015 & 0 & 21.651 & 0.038 &\\
\hline
\end{tabular}
}
\scalebox{0.97}{ 
 \begin{tabular}{rrrcrrcrrcccrcrrcrc}
\hline
\hline
&\multicolumn{3}{c}{$-Y-$}
&\multicolumn{3}{c}{$-J-$}
&\multicolumn{3}{c}{$-K_{\rm s}-$}
&\multicolumn{1}{l}{CLASS}
&\multicolumn{1}{c}{STAR}
&\multicolumn{2}{c}{DEIMOS}
&\multicolumn{3}{c}{Hectospec/HYDRA}
&\multicolumn{1}{c}{photoz}
&\multicolumn{1}{c}{SED $\chi^2$}
\\
&\multicolumn{1}{c}{mag}
&\multicolumn{1}{c}{err}
&\multicolumn{1}{c}{flag}
&\multicolumn{1}{c}{mag}
&\multicolumn{1}{c}{err}
&\multicolumn{1}{c}{flag}
&\multicolumn{1}{c}{mag}
&\multicolumn{1}{c}{err}
&\multicolumn{1}{c}{flag}
&\multicolumn{1}{r}{\_STAR}
&\multicolumn{1}{c}{}
&\multicolumn{1}{c}{$z_{\rm s}$}
&\multicolumn{1}{c}{flag}
&\multicolumn{1}{c}{$z_{\rm s}$}
&\multicolumn{1}{c}{flag}
&\multicolumn{1}{c}{class}
&\multicolumn{1}{c}{}
&\multicolumn{1}{c}{classification}
\\
&\multicolumn{1}{c}{(18)}
&\multicolumn{1}{c}{(19)}
&\multicolumn{1}{c}{(20)}
&\multicolumn{1}{c}{(21)}
&\multicolumn{1}{c}{(22)}
&\multicolumn{1}{c}{(23)}
&\multicolumn{1}{c}{(24)}
&\multicolumn{1}{c}{(25)}
&\multicolumn{1}{c}{(26)}
&\multicolumn{1}{c}{(27)}
&\multicolumn{1}{c}{(28)}
&\multicolumn{1}{c}{(29)}
&\multicolumn{1}{c}{(30)}
&\multicolumn{1}{c}{(31)}
&\multicolumn{1}{c}{(32)}
&\multicolumn{1}{c}{(33)}
&\multicolumn{1}{c}{(34)}
&\multicolumn{1}{c}{(35)}
\\
\hline
3 & -99.000 & -99.000 & -1 & -99.000 & -99.000 & -1 & -99.000 & -99.000 & -1 & 0.61 & **** & -99.000 & -99 & -99.000 & -99.0 & NA & 2.325&S\\
721 & -99.000 & -99.000 & -1 & -99.000 & -99.000 & -1 & -99.000 & -99.000 & -1 & 0.03 & **** & -99.000 & -99 & 0.215 & 4.0 & GALAXY & 0.275&G\\
5414 & -99.000 & -99.000 & -1 & -99.000 & -99.000 & -1 & -99.000 & -99.000 & -1 & 0.07 & **** & -99.000 & -99 & -99.000 & -99.0 & NA & 1.246&G\\
15159 & 21.044 & 0.106 & 0 & 21.013 & 0.115 & 0 & 21.201 & 0.171 & 0 & 0.93 & STAR & -99.000 & -99 & -99.000 & -99.0 & NA & 0.000&S\\
21777 & -99.000 & -99.000 & -1 & -99.000 & -99.000 & -1 & 22.064 & 0.093 & 0 & 0.59 & **** & -99.000 & -99 & -99.000 & -99.0 & NA & 0.259&G\\
44667 & 22.233 & 0.103 & 0 & 22.357 & 0.141 & 0 & 21.703 & 0.081 & 0 & 0.07 & **** & -99.000 & -99 & -99.000 & -99.0 & NA & 1.337&G\\
69588 & 20.967 & 0.030 & 0 & 20.599 & 0.044 & 0 & 20.129 & 0.034 & 0 & 0.07 & **** & 0.586 & 3 & -99.000 & -99.0 & NA & 0.655&G\\
79677 & 17.664 & 0.009 & 0 & 17.229 & 0.006 & 0 & 16.692 & 0.005 & 0 & 0.03 & **** & -99.000 & -99 & 0.028 & 2.0 & UNKNOWN & 0.253&G\\
84181 & -99.000 & -99.000 & -1 & -99.000 & -99.000 & -1 & -99.000 & -99.000 & -1 & 0.18 & **** & -99.000 & -99 & 1.081 & 3.0 & TYPE1 & 2.346&G\\
\hline
\end{tabular}
}
\label{tb:example}
\end{table}
\end{landscape}

\end{document}